\documentclass[a4paper,fleqn,usenatbib,useAMS]{mnras}
\usepackage{upgreek}
\usepackage{graphicx}
\usepackage{amsmath}
\usepackage{amssymb}
\usepackage{bm}
\usepackage{pdflscape}
\usepackage{float}
\usepackage[T1]{fontenc}
\usepackage{ae,aecompl}
\usepackage{color}
\usepackage{caption}
\usepackage[shortlabels]{enumitem}

\pdfoutput=1

\title[Tidal Evolution of Eccentric Binaries]{Tidal Evolution of Eccentric Binaries Driven by Convective Turbulent Viscosity}

\author[M. Vick and D. Lai]{Michelle Vick$^{1}$, Dong Lai$^{1}$\\$^{1}$Cornell Center for Astrophysics and Planetary Science, Department of Astronomy, Cornell University, Ithaca, NY 14853, USA}

\pubyear{2019}

\begin{document}
	
	
\label{firstpage}
\pagerange{\pageref{firstpage}--\pageref{lastpage}}
\maketitle
	
\begin{abstract}
Tidal dissipation due to convective turbulent viscosity shapes the evolution of a variety of astrophysical binaries. For example, this type of dissipation determines the rate of orbital circularization in a binary with a post-main sequence star that is evolving toward a common envelope phase. Viscous dissipation can also influence binaries with solar-type stars, or stars with a close-in giant planet. In general, the effective viscosity in a convective stellar envelope depends on the tidal forcing frequency $\omega_{\rm tide}$; when $\omega_{\rm tide}$ is larger than the turnover frequency of convective eddies, the viscosity is reduced.  Previous work has focused on binaries in nearly circular orbits. However, for eccentric orbits, the tidal potential has many forcing frequencies. In this paper, we develop a formalism for computing tidal dissipation that captures the effect of frequency-dependent turbulent viscosity and is valid for arbitrary binary eccentricities. We also present an alternative simpler formulation that is suitable for very high eccentricities. We apply our formalisms to a giant branch (GB) star model and a solar-type star model. We find that a range of pseudosynchronous rotation rates are possible for both stellar models, and the pseudosynchronous rate can differ from the prediction of the commonly-used weak tidal friction theory by up to a factor of a few. We also find that tidal decay and circularization due to turbulent viscosity can be a few orders of magnitude faster than predicted by weak tidal friction in GB stars on eccentric, small pericentre orbits, but is suppressed by a few orders of magnitude in solar-type stars due to viscosity reduction.
\end{abstract}
	

\section{Introduction}\label{sec:Intro}	
Tidal dissipation shapes a variety of astrophysical binaries, causing spin synchronization of the two bodies as well as orbital decay and circularization. For example, the evolution of a stellar binary toward a common envelope episode (CEE) is affected by tides. A CEE occurs when a binary system shares a gaseous envelope. The embedded system experiences drag forces that tighten the binary orbit \citep[e.g.][]{ Paczynski76,VDH76}. Many astronomical transients are believed to originate from systems that have experienced a CEE \citep{Belczynski02a,Dominik12,Belczynski18}. Some transients may be directly associated with a CEE, e.g., recent work has suggested that luminous red novae may be caused by the ejection of a common envelope \citep{Ivanova13,MacLeod17,Blagorodnova17}. A CEE can also account for the formation of compact double neutron star or black hole binaries, whose ultimate coalescence is detectable with LIGO/Virgo \citep[e.g.][]{Bhattacharya91,Belczynski02b,Tauris06,Dominik12,VG18}. In many cases, the onset and outcome of a CEE may depend on the orbital configuration of the binary when one component, a compact star, makes contact with the expanded convective envelope of the other component, a giant star. Because the pre-CEE binary can have a rather eccentric orbit, the strength of tidal dissipation in the convective envelope compared with the timescale for radius expansion of the giant star determines the orbit of the binary at the start of a CEE.

Tidal dissipation in a convective envelope can be also be important in binaries containing solar-type stars. Previous works have studied and compared different dissipation mechanisms within these stars. In the convective envelope, fundamental and inertial oscillation modes dissipate due to turbulent viscosity, while radiative diffusion operates in the stellar interior \citep[e.g.][]{Zahn77,Goodman97,Goodman98,Savonije02, Ogilvie07}. In some cases, internal gravity waves excited at the radiative-convective boundary grow in amplitude as they travel toward the center until non-linear wave-breaking dissipates the energy and angular momentum in the wave \citep{Goodman98,Barker10,Barker11,Chernov13,Ivanov13,Mathis16,Weinberg17,Sun18}. This effect can drive rapid orbital decay. However, when the internal gravity waves do not achieve nonlinearity, turbulent viscosity in the convective envelope is often the dominant mechanism of tidal dissipation.

A large body of work has studied the dissipation of equilibrium tides in the convective envelope of a star. An analytical treatment was first developed by \citet{Zahn77}, using an eddy viscosity $\nu_0 \sim v_H l/3$, where $l\sim H$ is the length-scale of the largest convective eddies and $v_H \sim (F/\rho)^{1/3}$ is the convective velocity on scale $H$ ($H$ is the pressure scale height, $F$ the convective flux, and $\rho$ the density). However, when the timescale of tidal forcing is shorter than the eddy turnover time, $\tau_{\rm eddy} = H/v_H$, convective eddies cannot efficiently transport energy and momentum. \citet{Zahn89} proposed that, in this case, the viscosity should be reduced by a linear factor of $(\omega \tau_{\rm eddy})^{-1}$, where $\omega$ is the tidal forcing frequency. In contrast, \citet{Goldreich77} suggested that the viscosity reduction should scale with $(\omega \tau_{\rm eddy})^{-2}$. Recently, numerical and analytical studies have supported a quadratic reduction factor \citep{Penev11a,Penev11b,Ogilvie12,Duguid19}. A few have even discovered negative viscosities at high forcing frequencies \citep{Ogilvie12,Duguid19}.

While many previous studies have explored the effect of convective viscous dissipation on binaries in circular orbits, very few have considered the effect of frequency dependent viscosity reduction for highly eccentric binaries. A star on an eccentric orbit experiences multiple tidal forcing frequencies; the more eccentric the orbit, the wider the tidal frequency spectrum becomes. \citet{IP04b} studied how different prescriptions for viscosity reduction affect the orbital evolution of a binary with a fully convective primary (e.g. a low-mass star or planet). They found that when the viscosity reduction scales more steeply than $(\omega \tau_{\rm eddy})^{-1}$, the orbital evolution of the system can change drastically.

In this paper, we study the effects of tidal dissipation due to convective turbulent viscosity in an eccentric binary. We consider both giant branch (GB) stars and solar-type stars, although our method can be applied to other types of stars with convective envelopes. In Section~\ref{sec:Theory}, we develop a general formalism for tidal evolution in an eccentric binary, accounting for frequency dependent damping of tidally excited oscillations. We relate this formalism to the standard weak friction treatment of the equilibrium tide \citep{Darwin1880,Alexander73,Hut81}. In Section~\ref{sec:Models} we introduce two stellar models (a GB model and a solar-type model) and discuss the effects of frequency dependent viscosity reduction in both. In Section~\ref{sec:Results} we present results for the tidal energy and angular momentum transfer rates for both stellar models as a function of the binary orbital parameters. In Section~\ref{sec:HighEcc} we discuss an alternative (and simpler) calculation of the transfer rates in the case of highly eccentric  ($\gtrsim0.8$) binaries before concluding in Section~\ref{sec:Discussion}.

\section{Tides and Dissipation in Eccentric Binaries: General Formalism}\label{sec:Theory}

	Consider a primary star with mass $M_1$ and radius $R_1$ in an orbit with semi-major axis $a$ and eccentricity $e$ about a secondary star, $M_2$. We study tidal dissipation in the convective envelope of the primary, neglecting the tide in the secondary. 
	
	In an inertial frame, the quadrupolar ($l=2$) tidal potential  produced by $M_2$ is
		\begin{equation}
		U(\boldsymbol{r}, t) = - GM_2\sum_{m}\frac{W_{2m}r^2}{D^{3}}\text{e}^{-\text{i}m\Phi(t)} Y_{2m}(\theta,\phi_i), \label{eq:potential}
		\end{equation}
	where $\boldsymbol{r} = (r,\theta,\phi_i=\phi+\Omega_s t)$ is the position vector in spherical coordinates relative to the center of mass of of the primary star, and the angle $\phi$ is measured in the rotating frame of $M_1$, which rotates with frequency $\Omega_s$. Throughout this paper, we assume that the spin axis of the star is aligned with the orbital angular momentum axis \cite[see, e.g.,][for the gravitational potential of $M_2$ if the spin axis and orbital axis are misaligned.]{Lai06} The time-varying binary separation is $D(t)$, and $\Phi(t)$ is the orbital true anomaly. Only the $m =  0, \pm2$ terms are nonzero, with $W_{20} = \sqrt{\uppi/5}$ and $W_{2\pm2} = \sqrt{3 \uppi/10}$. The potential can be decomposed into terms with frequencies that are integer multiples of the orbital frequency, $\Omega = [G(M_1+M_2)a^{-3}]^{1/2}$. In the rotating frame of the primary, we have
		\begin{equation}
		U(\boldsymbol{r}, t) = -  \sum_{m}\sum_{N=-\infty}^{\infty} U_{Nm} r^2 Y_{2m}(\theta,\phi)\text{e}^{-\text{i}\omega_{Nm}t},\label{eq:potentialSum}
		\end{equation} 
	where 
		\begin{align}
		U_{Nm} &\equiv \frac{GM_2}{a^3} W_{2m} F_{Nm},\\
		\omega_{Nm} &\equiv N\Omega - m \Omega_s, \label{eq:UNm}
		\end{align}
	with $F_{Nm}$ defined by the expansion
		\begin{equation}
		\left(\frac{a}{D}\right)^{3}\text{e}^{-\text{i}m \Phi(t)} = \sum_{N=-\infty}^{\infty}F_{Nm}\text{e}^{-\text{i} N \Omega t},
		\end{equation}
	and given by
		\begin{equation}
		F_{Nm} = \frac{1}{\uppi}\int_0^\uppi d\Psi \; \frac{\cos[N(\Psi - e \sin \Psi)-m\Phi(t)]}{(1-e\cos \Psi)^2},
		\end{equation}
	where $\Psi$ is the eccentric anomaly.
	
	The linear response of $M_1$ is specified by the Lagrangian displacement vector, $\bxi(\boldsymbol{r},t)$, which satisfies the equation of motion
		\begin{equation}
		\frac{\partial^2 \bxi}{\partial t^2} + 2 \boldsymbol{\Omega_s} \times \frac{\partial \bxi}{\partial t} + \boldsymbol{C} \cdot \bxi = - \nabla U,
		\end{equation}
	in the rotating frame of the primary, where $\boldsymbol{C}$ is a self-adjoint operator that contains the restoring forces acting on the perturbation. We can decompose the Lagrangian displacement into a sum of eigenmodes $\bxi_\alpha(\boldsymbol{r})$ of frequency $\omega_\alpha$ (where $\alpha$ specifies the mode indices, which include the degree, $l$, and azimuthal index, $m$) such that \citep{Schenk02,Lai06}
		\begin{equation}
		\begin{bmatrix}
		\bxi \\ \partial \bxi/\partial t
		\end{bmatrix}
		 = \sum_{\alpha} c_\alpha(t)
		 \begin{bmatrix}
		 \bxi_\alpha (\boldsymbol{r})\\
		 - i \omega_\alpha \bxi_\alpha (\boldsymbol{r})
		 \end{bmatrix},\label{eq:modedef}
		\end{equation}
	where $\omega_\alpha$ is the mode frequency in the rotating frame. Note that the above decomposition includes both positive and negative mode frequencies. We adopt the convention that the eigenmode oscillation has the form $\bxi(\boldsymbol{r},t)\propto\text{e}^{\text{i}m\phi - \text{i} \omega_\alpha t}$ such that positive $\omega_\alpha/m$ corresponds to a prograde mode. We use the normalization 
		\begin{equation}
		\langle \bxi_\alpha, \bxi_{\alpha} \rangle
		\equiv \int d^3x \rho \bxi^*_\alpha \cdot \bxi_\alpha = M_1 R_1^2,
		\end{equation}
	where $\rho$ is the stellar density profile. With this phase space expansion, the modes satisfy the orthogonality relation $\langle \bxi_\alpha, 2i \boldsymbol{\Omega_s} \times \bxi_{\beta}\rangle + (\omega_\alpha + \omega_{\beta} )\langle \bxi_\alpha, \bxi_{\beta}\rangle = 0$ for $\alpha \ne \beta$. We define 
		\begin{equation}
		c_\alpha(t) = \sum_{N=-\infty}^{\infty} c_{\alpha N}(t), \label{eq:calphaN_def}
		\end{equation}
	and find \citep{Lai06,Fuller12a}
		\begin{equation}
		\dot{c}_{\alpha N} + \left[i\omega_\alpha + \Gamma_{\alpha}(\omega_{Nm})\right]c_{\alpha N} = i \frac{U_{Nm}Q_\alpha}{2 \epsilon_\alpha}\text{e}^{-\text{i}\omega_{Nm}t}, \label{eq:cdot}
		\end{equation}
	where we have used
		\begin{align}
		Q_\alpha &= \langle \bxi_\alpha, \nabla (r^2 Y_{lm})\rangle, \label{eq:defQ}\\
		\epsilon_\alpha &= \omega_\alpha + \langle \bxi_\alpha, \text{i} \boldsymbol{\Omega_s}\times \bxi_\alpha\rangle. \label{eq:defepsilon}
		\end{align}
	To the first order in the stellar rotation rate, $\epsilon_\alpha$ is the eigenfrequency of a mode in the absence of rotation, provided $\Omega_s \ll |\omega_\alpha|$. The damping rate of the forced oscillation of mode $\alpha$ at the forcing frequency $\omega_{Nm}$ is denoted by $\Gamma_{\alpha}(\omega_{Nm})$. The solution to equation~(\ref{eq:cdot}) is
	\begin{align}
	c_{\alpha N}(t) = \frac{U_{Nm} Q_{\alpha}}{2 \epsilon_\alpha} \frac{\text{e}^{-\text{i} \omega_{Nm} t}}{[\omega_\alpha - \omega_{Nm}-\text{i} \Gamma_{\alpha}(\omega_{Nm})]} \label{eq:c_solution},
	\end{align}
	
	On timescales much longer than the orbital period, the energy dissipation rate in the rotating frame is given by the sum over the response to multiple forcing frequencies for oscillation modes, i.e,
	\begin{equation}
		\dot{E} = \sum_{\alpha N}\dot{E}_{\alpha N}.
	\end{equation}
	In this paper, we consider tidal dissipation due to viscosity in the stellar convection zone (see Section~\ref{sec:Models}). For slow rotation ($\Omega_s \ll|\omega_\alpha|$), the oscillation eigenmode is given by 
		\begin{equation}
		\bxi_\alpha(\boldsymbol{r}) = \xi_{\alpha r}(\boldsymbol{r})Y_{lm}(\theta,\phi)\boldsymbol{e}_r + \xi_{\alpha h}(\boldsymbol{r}) r\boldsymbol{\nabla} Y_{lm}(\theta,\phi),
		\end{equation}
	where $\xi_{\alpha r}$ and $\xi_{\alpha h}$ are the radial and horizontal components of $\bxi_\alpha({\textbf{r}})$. The viscous dissipation rate of mode $\alpha$, oscillating at the forcing frequency $\omega_{Nm}$ with amplitude $c_{\alpha N}$ is [see equation~(5) of \citet{Sun18}]
		\begin{align}
		\dot{E}_{\alpha N} =& \frac{1}{2} \omega_{Nm}^2 |c_{\alpha N}|^2 \nonumber \\
		&\times \int_{r_{\rm conv}}^{R_1} dr \;r^2 \rho \nu \left[4 \left(\frac{d\xi_{\alpha r}}{dr}\right)^2 \right. \nonumber\\
		&+ \left. 2l(l+1)\left(\frac{d \xi_{\alpha h}}{dr} + \frac{\xi_{\alpha r}}{r}-\frac{\xi_{\alpha h}}{r}\right)^2 \right. \nonumber \\
		&+ \left. 2\left(l(l+1)\frac{\xi_{\alpha h}}{r} - 2\frac{\xi_{\alpha r}}{r}\right)^2\right], \label{eq:EdotAlphaN}
		\end{align}
	where $r_{\rm conv}$ is the inner edge of the convective envelope, and $\nu$ is the isotropic kinematic viscosity. Equation~(\ref{eq:EdotAlphaN}) assumes that the flow is approximately incompressible. We define the damping rate $\gamma_\alpha(\omega_{Nm})$ as a relationship between $\dot{E}_{\alpha N}$ and the kinetic energy of the mode such that 
	\begin{align}
		\dot{E}_{\alpha N}  &= 2 \gamma_{\alpha} (\omega_{Nm}) \langle \dot{\bxi}_{\alpha N}, \dot{\bxi}_{\alpha N}\rangle \nonumber\\ 
		&= 2 \gamma_{\alpha}(\omega_{Nm}) \omega_{Nm}^2 |c_{\alpha N}|^2 M_1 R_1^2,
	\end{align}
	with $\bxi_{\alpha N} (\boldsymbol{r},t)\equiv c_{\alpha N} (t)  \bxi_\alpha (\boldsymbol{r})$. Thus, 
	\begin{align}
	\gamma_\alpha (\omega_{Nm}) \equiv& \frac{1}{4}\int_{r_{\rm conv}}^{R_1} dr \;r^2 \rho \nu \left[4 \left(\frac{d\xi_{\alpha r}}{dr}\right)^2 \right. \nonumber \\
	&+ \left. 2l(l+1)\left(\frac{d \xi_{\alpha h}}{dr} + \frac{\xi_{\alpha r}}{r}-\frac{\xi_{\alpha h}}{r}\right)^2 \right. \nonumber \\
	&+ \left. 2\left(l(l+1)\frac{\xi_{\alpha h}}{r} - 2\frac{\xi_{\alpha r}}{r}\right)^2\right]. \label{eq:defgamma}
	\end{align}
	The relationship between the two damping rates, $\Gamma_\alpha(\omega_{Nm}) = \omega_{Nm}\gamma_\alpha(\omega_{Nm})/\epsilon_\alpha$, is discussed in the Appendix. The total energy dissipation rate in the rotating frame is then given by the sum
		\begin{equation}
		\dot{E} =\frac{GM_2^2 R_1^5}{2a^6}\sum_{\alpha N} \left(\frac{Q_\alpha}{\bar{\epsilon}_\alpha}\right)^2 \frac{\gamma_{\alpha}(\omega_{Nm}) (W_{2m}F_{Nm})^2 \omega_{Nm}^2}{(\omega_\alpha - \omega_{Nm})^2 + \Gamma_{\alpha}^2(\omega_{Nm})} \label{eq:Edot}.
		\end{equation}
	In the above expression, $Q_\alpha$ and $\bar{\epsilon}_\alpha = \epsilon_\alpha(GM_1/R_1^3)^{-1/2}$ are dimensionless (i.e. they are in units where $G=M_1=R_1=1$). Note that if we restrict to modes with positive $\omega_\alpha$ in the sum, we can combine terms with $\omega_{\alpha}, m, N$  and $-\omega_{-\alpha},-m,-N$. The result is to multiply equation~(\ref{eq:Edot}) by a factor of 2. We can calculate the tidal torque on the primary, $T = \sum_{\alpha N} T_{\alpha N}$ using the relationship 
		\begin{equation}
		T_{\alpha N} = \frac{m\dot{E}_{\alpha N}}{\omega_{Nm}}. \label{eq:Ldot}
		\end{equation}
	Then,
		\begin{equation}
		T = \sum_{\alpha N} T_{\alpha N} = T_0\sum_{\alpha' N} \left(\frac{Q_\alpha}{\bar{\epsilon}_\alpha}\right)^2 \frac{m \gamma_{\alpha}(\omega_{Nm}) (W_{2m}F_{Nm})^2 \omega_{Nm}}{(\omega_\alpha - \omega_{Nm})^2 + \Gamma_{\alpha}^2(\omega_{Nm})} \label{eq:Tdef},
		\end{equation}
	where
		\begin{equation}
		T_0 \equiv \frac{GM_2^2R_1^5}{a^6}, \label{eq:T0def}
		\end{equation}
	and $\sum_{\alpha'}$ implies that the sum is restricted to modes with $\omega_\alpha>0$.
	The tidal energy transfer rate from the orbit to the primary in the inertial frame, $\dot{E}_{\rm in}$, is related to $\dot{E}$ and $T$ via
		\begin{equation}
		\dot{E}_{\rm in} = \dot{E} + \Omega_s T.\label{eq:Edotin_and_Edot}
		\end{equation}
	From equations~(\ref{eq:Edot})-(\ref{eq:Edotin_and_Edot}), we find
		\begin{align}
		\dot{E}_{\rm in} &= T_0 \Omega \sum_{\alpha' N} \left(\frac{Q_\alpha}{\bar{\epsilon}_\alpha}\right)^2 \frac{N\gamma_{\alpha}(\omega_{Nm}) (W_{2m}F_{Nm})^2 \omega_{Nm}}{(\omega_\alpha - \omega_{Nm})^2 + \Gamma_{\alpha}^2(\omega_{Nm})}. \label{eq:Edotin}
		\end{align}
	Together, equations~(\ref{eq:Tdef}) and (\ref{eq:Edotin}) govern the spin and orbital evolution of the binary (see Section~\ref{sec:EqSummary}). \footnote{The derivation of equations~(\ref{eq:Tdef}) and (\ref{eq:Edotin}) for the tidal torque and energy transfer rate differs from \citet{IP04b} in that we use a mode decomposition (equation~\ref{eq:modedef}) that is rigorously valid for rotating stars.}

	\subsection{The Slow-Rotation and Weak Friction Limits}
	In the limit $\Omega_s \ll \omega_\alpha$, $\omega_\alpha \simeq \omega_{-\alpha} \simeq \epsilon_\alpha \simeq \epsilon_{-\alpha}$. Assuming $\omega_\alpha \gg|\omega_{Nm}|$ and $\omega_\alpha \gg \Gamma_\alpha(\omega_{Nm})$, equations~(\ref{eq:Tdef}) and (\ref{eq:Edotin}) can be simplified to 
	\begin{align}
	T &\simeq T_0 \sum_{\alpha' N} \left(\frac{Q_\alpha}{\bar{\omega}_\alpha}\right)^2 \frac{m\gamma_{\alpha}(\omega_{Nm}) (W_{2m}F_{Nm})^2 \omega_{Nm}}{\omega_\alpha^2} \label{eq:T_slowRot},\\
	\dot{E}_{\rm in} &\simeq T_0 \Omega \sum_{\alpha' N} \left(\frac{Q_\alpha}{\bar{\omega}_\alpha}\right)^2 \frac{N\gamma_{\alpha}(\omega_{Nm}) (W_{2m}F_{Nm})^2 \omega_{Nm}}{\omega_\alpha^2} \label{eq:Edot_slowRot}.
	\end{align}
	
	In the weak friction approximation \citep{Darwin1880,Alexander73,Hut81}, the damping rate $\gamma_\alpha(\omega_{Nm})= \gamma_\alpha$ is assumed to have no frequency dependence, and we expect equations~(\ref{eq:T_slowRot}) and (\ref{eq:Edot_slowRot}) to reduce to the standard result from, e.g. \citet{Alexander73,Hut81}. To see this, we first identify the tidal Love number and lag time. The complex Love number associated with each forcing term ($Nm$) is
	\begin{equation}
	\tilde{k}_2(\omega_{Nm}) = \left.\frac{\left[\delta \Phi(\boldsymbol{r},t)\right]_{Nm}}{[U(\boldsymbol{r},t)]_{Nm}}\right\vert_{r=R_1},
	\end{equation}
	where $\delta \Phi(\boldsymbol{r},t)$ is the potential from the perturbed density in the primary star. Using,
	\begin{equation}
	\left[\delta \Phi(\boldsymbol{r},t)\right]_{Nm} = \sum_{\alpha} c_{\alpha N}(t) \delta \Phi_\alpha(\boldsymbol{r}),
	\end{equation}
	with the expansion in spherical harmonics (limited to $l=2$)
	\begin{equation}
	\left. \delta \Phi_\alpha(\boldsymbol{r}) \right\vert_{r=R_1} = -\frac{4 \pi}{5}\frac{GM_1}{R_1}Q_\alpha Y_{2m}(\theta,\phi), \label{eq:deltaPhi}
	\end{equation}
	we find that, for an f-mode oscillation in a slowly rotating body, the real part of the tidal Love number is
	\begin{equation}
	k_2 \simeq \frac{4 \pi}{5}\left(\frac{Q_{\rm f}}{\bar{\omega}_{\rm f}}\right)^2, \label{eq:k2def}
	\end{equation}
	where $Q_{\rm f}$ and $\bar{\omega}_{\rm f} = \omega_{\rm f}(GM_1/R_1^3)^{-1/2}$ are the overlap integral and eigenfrequency of the $l=2$ f-mode for a non-rotating body. The tidal lag time can be defined as
	\begin{equation}
	\tau \equiv \frac{\gamma_{\rm f}}{\omega_{\rm f}^2},\label{eq:taudef}
	\end{equation}
	where $\gamma_{\rm f}$ is the damping rate of the $l=2$ f-mode oscillation calculated with equation~(\ref{eq:defgamma}) assuming that the kinematic viscosity is independent of the forcing frequency. The sum over oscillation modes in equations~(\ref{eq:Tdef}) and (\ref{eq:Edotin}) is restricted to f-modes with $m=-2,0,2$. The tidal torque and energy transfer rate can then be written as
	\begin{align}
	T &= 3 T_0 k_2 \tau \Omega \sum_{Nm}\frac{5}{12\pi} m  (W_{2m}F_{Nm})^2 \frac{\omega_{Nm}}{\Omega} \label{eq:T_WF},\\
	\dot{E}_{\rm in} &= 3  T_0 k_2 \tau \Omega^2\sum_{Nm} \frac{5}{12\pi} N (W_{2m}F_{Nm})^2 \frac{\omega_{Nm}}{\Omega} \label{eq:Edot_WF}.
	\end{align}
	According to equations~(22) and (23) of \citet{Storch14}, we have
	\begin{align}
	\sum_{Nm}\frac{5}{12\pi} m  (W_{2m}F_{Nm})^2 &\frac{\omega_{Nm}}{\Omega} = \frac{1}{(1-e^2)^6} \nonumber \\
	&\times \left[f_2 - (1-e^2)^{3/2}f_5\frac{\Omega_s}{\Omega}\right] \\
	\sum_{Nm}\frac{5}{12\pi} N  (W_{2m}F_{Nm})^2 &\frac{\omega_{Nm}}{\Omega} = \frac{1}{(1-e^2)^{15/2}} \nonumber \\
	&\times \left[f_1 - (1-e^2)^{3/2}f_2\frac{\Omega_s}{\Omega}\right],
	\end{align}
	where $f_1$, $f_2$, and $f_5$ are functions of eccentricity defined in \citet{Hut81}, given by
	\begin{align}
	f_1 &= 1+\frac{31}{2}e^2 + \frac{255}{8}e^4 + \frac{185}{16}e^6 + \frac{25}{64}e^8, \\
	f_2 &= 1 + \frac{15}{2}e^2 + \frac{45}{8}e^4 + \frac{5}{16}e^6, \\
	f_5 &= 1 + 3e^2 + \frac{3}{8}e^4.
	\end{align}
	This verifies that our formulation is equivalent to the weak friction model under the assumptions of slow rotation and frequency-independent viscous dissipation.
	
\subsection{Orbital Evolution} \label{sec:EqSummary}
	
	We can combine the angular momentum and energy transfer rates to obtain the orbital evolution of the binary and spin evolution of the primary star driven by tidal dissipation. The rate of change of the orbital angular momentum is $\dot{L} = - T$, and the orbital energy dissipation rate is $\dot{E}_{\rm orb} = - \dot{E}_{\rm in}$. Using $L = \mu \Omega a^2 (1-e^2)^{1/2}$ and $E_{\rm orb} = - \mu \Omega^2 a^2 / 2$ (with $\mu$ the reduced mass of the binary), we find that 
	\begin{align}
	\frac{\dot{a}}{a} &= -\frac{2}{\mu \Omega^2 a^2}\dot{E}_{\rm in}, \label{eq:adot}\\
	\frac{\dot{\Omega}_s}{\Omega_s} &= \frac{T}{I_1\Omega_s}, \label{eq:OSdot}\\
	\frac{\dot{e}}{e} &= \frac{1-e^2}{e^2}\frac{1}{\mu \Omega a^2}\left[\frac{T}{(1-e^2)^{1/2}} - \frac{\dot{E}_{\rm in}}{\Omega}\right],\label{eq:edot}
	\end{align}
	where $I_1=kM_1R_1^2$ is the moment of inertia of the primary star. 
	
	To facilitate applications to different binary systems, we write $T$ and $\dot{E}_{\rm in}$ in the form 
	\begin{align}
	T &= 3 T_0 k_2\tau\frac{\Omega}{(1-e^2)^6}F_{T}(e,\Omega_s/\Omega,r_p/R_1), \label{eq:TGeneral}\\ 
	\dot{E}_{\rm in} &= 3 T_0 k_2\tau\frac{\Omega^2}{(1-e^2)^{15/2}}F_{E}(e,\Omega_s/\Omega,r_p/R_1), \label{eq:EdotGeneral}
	\end{align}
	where $k_2$ and $\tau$ are given by equations~(\ref{eq:k2def}) and (\ref{eq:taudef}), $r_p$ is the pericentre distance, and $F_T$ and $F_E$ are dimensionless functions that depend on $e$, $\Omega_s/\Omega$, $r_p/R$, and the structure of the star. Note that the semi-major axis evolution depends on $F_{E}$ and the stellar spin on $F_T$. We also define the quantity 
	\begin{equation}
	F_{\rm ecc}(e,\Omega_s/\Omega,r_p/R_1) = \frac{1}{9}\frac{(1-e^2)}{e^2}\left[\frac{F_E}{(1-e^2)} - F_T\right],
	\end{equation}
	which characterizes the eccentricity evolution. In the weak friction limit,
	\begin{align}
	F_T &= f_2 - (1-e^2)^{3/2}f_5\frac{\Omega_s}{\Omega}, \label{eq:FT_WF} \\
	F_E &= f_1 - (1-e^2)^{3/2}f_2\frac{\Omega_s}{\Omega} \label{eq:FE_WF} \\
	F_{\rm ecc} &= f_3-(1-e^2)^{3/2}f_4\frac{11}{18}\frac{\Omega_s}{\Omega},\label{eq:FEcc_WF}
	\end{align}
	where $f_3$ and $f_4$ are functions of the eccentricity defined in \citet{Hut81} and given by
	\begin{align}
	f_3 &= 1+ \frac{15}{4}e^2 + \frac{15}{8}e^4 + \frac{5}{64}e^6 \\
	f_4 &= 1 +\frac{3}{2}e^2 + \frac{1}{8}e^4.
	\end{align}
	
	For general binary systems (when the weak friction theory breaks down), we continue to use equations~(\ref{eq:TGeneral}) and (\ref{eq:EdotGeneral}) to parameterize the angular momentum and energy transfer rates. By comparing equations~(\ref{eq:Tdef}) and (\ref{eq:Edotin}) to equations~(\ref{eq:TGeneral}) and (\ref{eq:EdotGeneral}), we find
	\begin{align}
	F_T(e,&\Omega_s/\Omega,r_p/R_1) = \frac{5}{12\pi}\left(\frac{\bar{\omega}_{\rm f}}{Q_{\rm f}}\right)^{2}\left(\frac{\omega_{\rm f}^2}{\gamma_{\rm f} \Omega}\right)(1-e^2)^6  \nonumber \\
	&\times \sum_{Nm}\left(\frac{Q_\alpha}{\bar{\epsilon}_\alpha}\right)^2 \frac{m(W_{2m}F_{Nm})^2\omega_{Nm}\gamma_\alpha(\omega_{Nm})}{(\omega_\alpha-\omega_{Nm})^2+\Gamma_{\alpha}^2(\omega_{Nm})},  \label{eq:FT}\\
	F_E(e,&\Omega_s/\Omega,r_p/R_1) = \frac{5}{12\pi}\left(\frac{\bar{\omega}_{\rm f}}{Q_{\rm f}}\right)^{2}\left(\frac{\omega_{\rm f}^2}{\gamma_{\rm f} \Omega}\right)(1-e^2)^{15/2} \nonumber \\
	&\times \sum_{Nm} \left(\frac{Q_\alpha}{\bar{\epsilon}_\alpha}\right)^2 \frac{N(W_{2m}F_{Nm})^2\omega_{Nm}\gamma_\alpha(\omega_{Nm})}{(\omega_\alpha-\omega_{Nm})^2+\Gamma_{\alpha}^2(\omega_{Nm})}, \label{eq:FE}\\
	F_{\rm ecc}(e,&\Omega_s/\Omega,r_p/R_1) = \frac{5}{108 \pi}\left(\frac{\bar{\omega}_{\rm f}}{Q_{\rm f}}\right)^{2}\left(\frac{\omega_{\rm f}^2}{\gamma_{\rm f} \Omega}\right)\frac{(1-e^2)^{15/2}}{e^2} \nonumber \\
	&\times \sum_{Nm}\left[\left(N - \frac{m}{\sqrt{1-e^2}}\right)\left(\frac{Q_\alpha}{\bar{\epsilon}_\alpha}\right)^2 \right. \nonumber \\
	&\times \left.\frac{(W_{2m}F_{Nm})^2\omega_{Nm}\gamma_\alpha(\omega_{Nm})}{(\omega_\alpha-\omega_{Nm})^2+\Gamma_{\alpha}^2(\omega_{Nm})}\right]. \label{eq:FEcc}
	\end{align}
	We can now write the the orbital and spin evolution rates as
	\begin{align}
	\frac{\dot{a}}{a} &= -\frac{6}{t_d(1-e^2)^{15/2}}F_E(e,\Omega_s/\Omega,r_p/R_1),\label{eq:Simple_adot}\\
	\frac{\dot{\Omega}_s}{\Omega_s} &= \frac{3}{t_d(1-e^2)^{6}}\left(\frac{\mu a^2 \Omega}{I_1 \Omega_s}\right)F_T(e,\Omega_s/\Omega,r_p/R_1),\\
	\frac{\dot{e}}{e} &= -\frac{27}{t_d(1-e^2)^{13/2}} F_{\rm ecc}(e,\Omega_s/\Omega,r_p/R_1),\label{eq:Simple_edot}
	\end{align}
	where
	\begin{equation}
	t_d^{-1} \equiv \frac{T_0}{\mu a^2} k_2 \tau = \left(\frac{M_2}{M_1}\right)\left(\frac{M_1+M_2}{M_1}\right)\left(\frac{R_1}{a}\right)^8 k_2\frac{\gamma_{\rm f}}{\bar{\omega}_{\rm f}^2}. \label{eq:td_def}
	\end{equation}

	\section{Viscous Dissipation in Convective Envelopes and Stellar Models}\label{sec:Models}
	\begin{figure*}
		\begin{center}
			\includegraphics[width = 3.2 in]{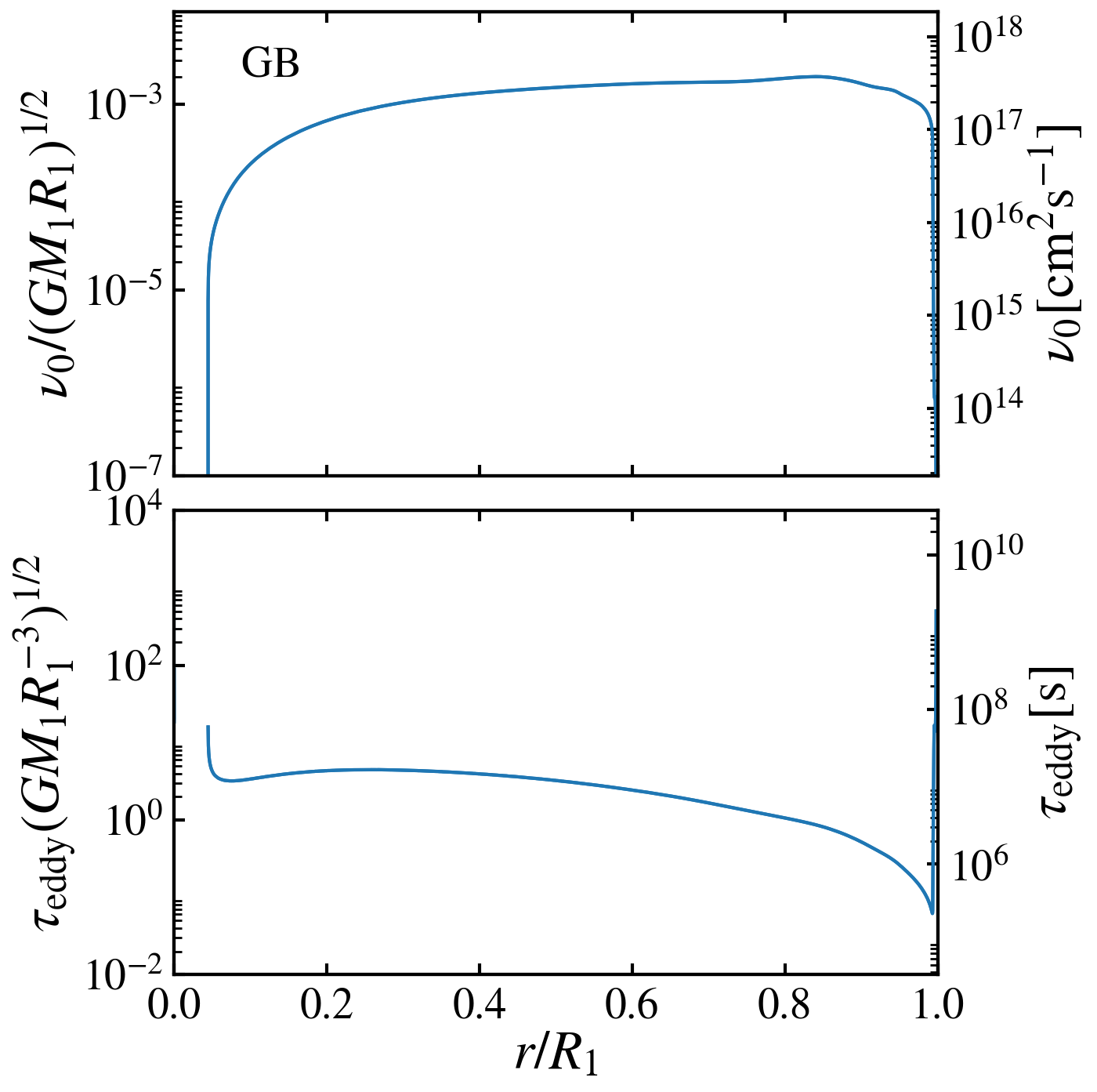}
			\includegraphics[width= 3.2 in]{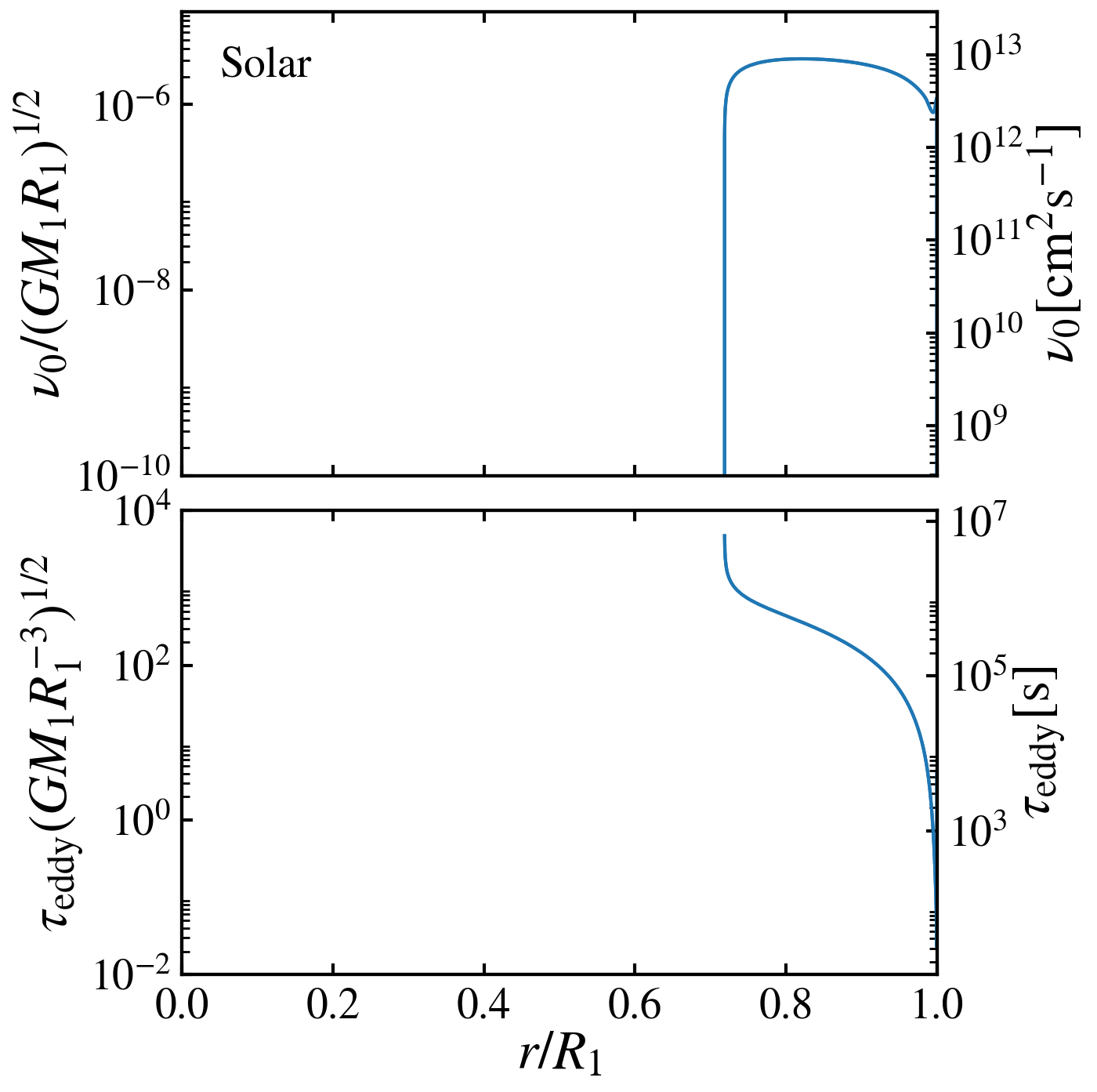}
			\caption{The standard (``unsuppressed") turbulent viscosity $\nu_0$ (top panels, see equation~\ref{eq:standard}) and eddy turnover time $\tau_{\rm eddy} = H/v_H$ (bottom panels) in the convective envelopes of two MESA-generated stellar models. The left panels show a giant branch star with $M_1 = 10 M_{\odot}$ and $R_1 = 379 R_{\odot}$, and the right panels show a solar-type star. In each panel, the left label shows $\nu_0$ and $\tau_{\rm eddy}$ in dimensionless ``stellar" units, while the right label shows the quantities in cgs units.}
			\label{fig:StellarModels}
		\end{center}
	\end{figure*}
	The damping rate of a forced oscillation mode depends on the convective viscosity. The standard viscosity prescription is independent of the forcing frequency and given by
	\begin{equation}
	\nu_0 = \frac{1}{3}H v_{H},\label{eq:standard}
	\end{equation}
	where $H$ is the pressure scale height, and $v_H \sim (F/\rho)^{1/3}$ is the convective velocity with $F$ the convective flux. In Fig.~\ref{fig:StellarModels}, we show the viscosity and eddy turnover time $\tau_{\rm eddy}= H/v_H$ for two MESA-generated stellar models \citep{Paxton11}, a $10 M_\odot$ giant branch (GB) star  and a solar-type star. 
	
	When the eddy turnover time  exceeds the tidal forcing period, $\sim |\omega_{Nm}|^{-1}$, convective eddies cannot transport momentum efficiently, and the turbulent viscosity is expected to be reduced. The correct prescription for viscosity reduction has been widely discussed in the literature \citep[see][for a review]{Ogilvie14}. We adopt a quadratic reduction, first suggested by \citet{Goldreich89} and confirmed in many recent studies [see e.g. \citet{Duguid19}],
	\begin{equation}
	\nu = \frac{1}{3}H v_{H}[1 + (\omega_{Nm}\tau_{\rm eddy})^2]^{-1}. \label{eq:GN}
	\end{equation}
	
\begin{figure}
		\begin{center}
			\includegraphics[width = 3.2 in]{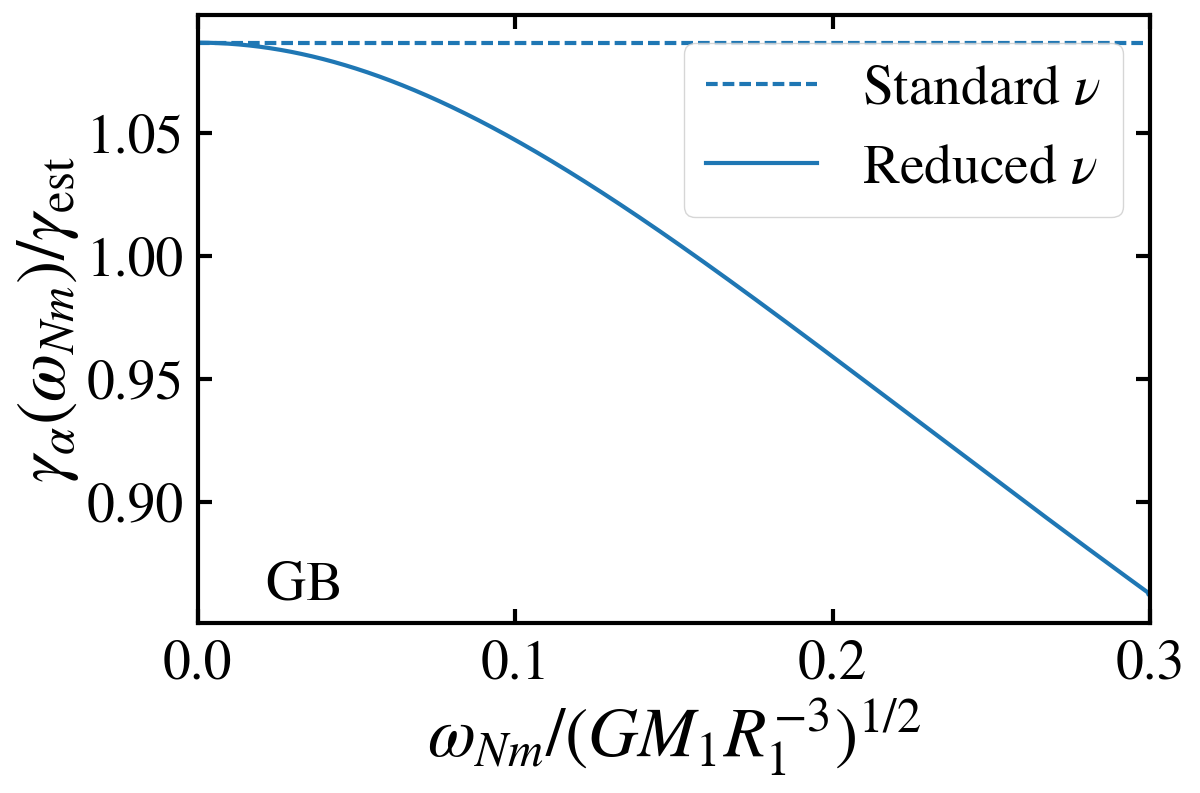}
			\includegraphics[width = 3.2 in]{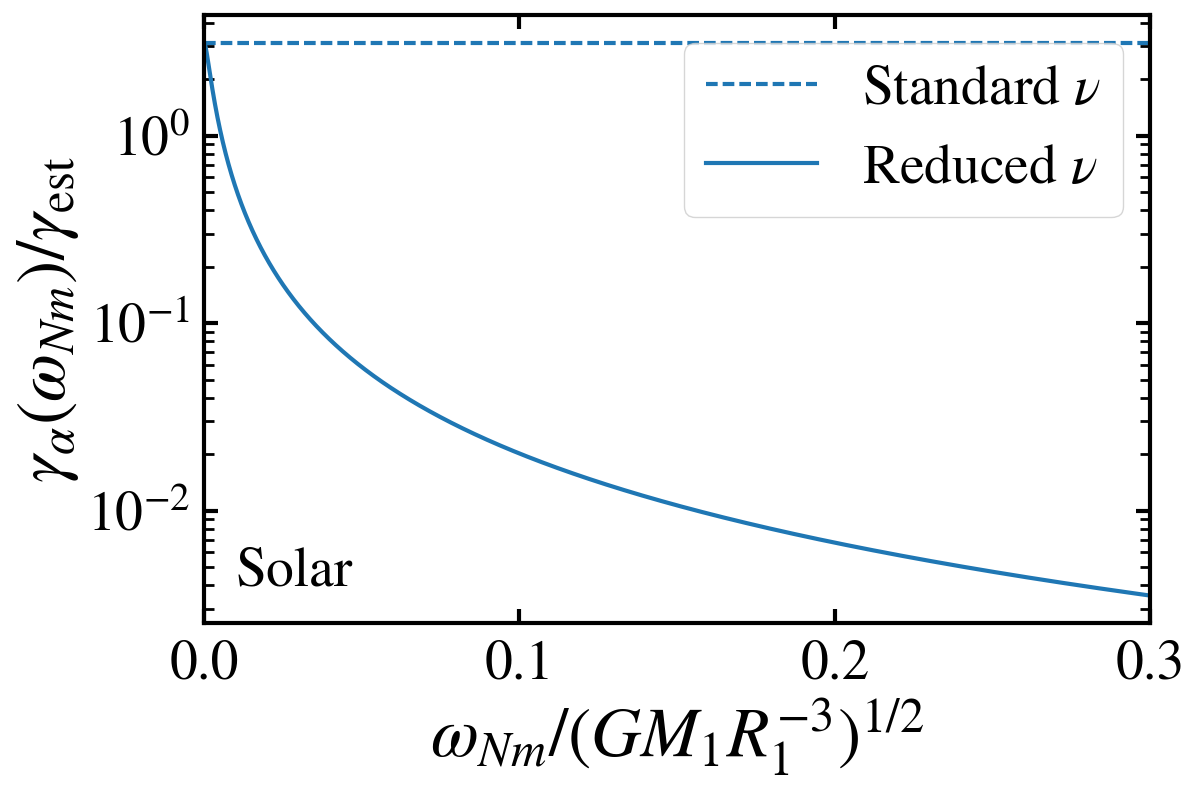}
			\caption{The stellar viscous damping rate (see equation~\ref{eq:defgamma}) as a function of the forcing frequency $\omega_{Nm}$ for the same MESA GB model (top panel) and solar model (bottom panel) used in Fig~\ref{fig:StellarModels}. The damping rates are scaled to $\gamma_{\rm est}$ (see equation~\ref{eq:gamma_est}). The dashed lines show the results for a standard (``unsuppressed") viscosity $\nu_0$ (equation~\ref{eq:standard}), and the solid line is calculated with a quadratic viscosity reduction (see equation~\ref{eq:GN}).}
			\label{fig:gammavsforcing}
		\end{center}
	\end{figure}

	\citet{Zahn77}, \citet{Phinney92} and \citet{Verbunt95} provided a simple estimate of the eccentricity circularization rate for nearly circular binary stars with convective envelopes based on the ``unsuppressed" viscosity $\nu_0$ (equation~\ref{eq:standard}). To estimate the damping rate of the tidally forced f-mode (i.e. equilibrium tide), we assume $\nu_0\sim$ constant in the convective envelope. Then the integral in equation~(\ref{eq:defgamma}) can be approximated as
	\begin{equation}
	\gamma_{\rm est} \sim  \frac{M_{\rm env}}{M_1}\left(\frac{\nu}{H^2}\right) \sim \frac{M_{\rm env}}{M_1}\left(\frac{L}{M_{\rm env}R_1^2}\right)^{1/3}, \label{eq:gamma_est}
	\end{equation}
	where $M_{\rm env}$ is the mass of the convective region and we have used $\nu_0 \sim H(F/\rho)^{1/3}$ and $4\pi \rho H^3 \sim M_{\rm env}$. Using this estimate of the equilibrium tide dissipation rate, \citet{Phinney92} gave the following approximation to the eccentricity damping rate (see Eq.~\ref{eq:td_def})
	\begin{equation}
	\frac{\dot{e}}{e} \approx -\gamma_{\rm est} \left(\frac{M_2}{M_1}\right)\left(\frac{M_1+M_2}{M_1}\right)\left(\frac{R_1}{a}\right)^8. \label{eq:Phinney}
	\end{equation}
	In Fig.~\ref{fig:gammavsforcing}, we compare $\gamma(\omega_{Nm})$ calculated with equation~(\ref{eq:defgamma}) to $\gamma_{\rm est}$ for the MESA-generated GB and solar-type stellar models used to produce Fig.~\ref{fig:StellarModels}. We find that $\gamma(\omega_{Nm} = 0) \sim \gamma_{\rm est}$. In stellar units, $\gamma_{\rm est} = 0.024 (GM_1/R_1^3)^{1/2}$ for the GB model and $\gamma_{\rm est} = 9.5\times 10^{-6} (GM_1/R_1^3)^{1/2}$ for the solar-type model, corresponding to $\gamma_{\rm est} = 0.20~\text{yr}^{-1}$ and $\gamma_{\rm est} = 0.22~\text{yr}^{-1}$ respectively. The eddy turnover time is generally larger in the solar-type model, so it is easier for the tidal forcing period to be comparable to $\tau_{\rm eddy}$ in the depths of the convective envelope. In consequence, viscosity reduction can have a very large effect on $\gamma(\omega_{Nm})$ for the solar model, depending on $\omega_{Nm}$, and is less important for the GB model.

	\section{Sample Results}\label{sec:Results}
		We have calculated $F_T(e,\Omega_s/\Omega,r_p/R_1)$ and $F_E(e,\Omega_s/\Omega,r_p/R_1)$ (defined in equations~\ref{eq:TGeneral} and \ref{eq:EdotGeneral}, see also equations~\ref{eq:FT}-\ref{eq:FEcc}) for the two stellar models introduced in Section~\ref{sec:Models}. These dimensionless angular momentum and energy transfer rates control the synchronization rate of the primary star and the orbital evolution of the binary (see equations~\ref{eq:Simple_adot} - \ref{eq:Simple_edot}). 
		
		Typically, the timescale for the primary star to reach an equilibrium spin rate, or pseudosynchronous rate, is shorter than the timescale for orbital decay and circularization. This is clear from equations~(\ref{eq:Simple_adot}) - (\ref{eq:Simple_edot}), where the spin evolution rate is faster than the orbital decay rate by a factor of order $\sim \mu a^2 \Omega/ I \Omega_s$. For an eccentric orbit, the primary star can spin-up to pseudosynchronous rotation very quickly, and it is safe to assume that the star is rotating pseudosynchronously throughout orbital decay and circularization.
		
		When a star rotates pseudosynchrounously, it experiences no net torque. Under the weak friction approximation,  the dimensionless torque is given by equation~(\ref{eq:FT_WF}), so the pseudosynchronous rate is 
		\begin{equation}
		\Omega_{\rm ps} = \frac{f_2}{(1-e^2)^{3/2}f_5}\Omega = \frac{f_2}{(1+e)^2 f_5}\Omega_{\rm p},\label{eq:WFOS}
		\end{equation}
	where 
		\begin{equation}
		\Omega_{\rm p} \equiv \left[\frac{(1+e)G(M_1+M_2)}{r_p^3}\right]^{1/2} = \Omega \frac{(1+e)^{1/2}}{(1-e)^{3/2}}, \label{eq:Omegap_def}
		\end{equation}
	is the orbit frequency at pericenter [$r_{\rm p} = (1-e)a$]. 
	
\begin{figure*}
			\begin{center}
				\includegraphics[width = 5 in]{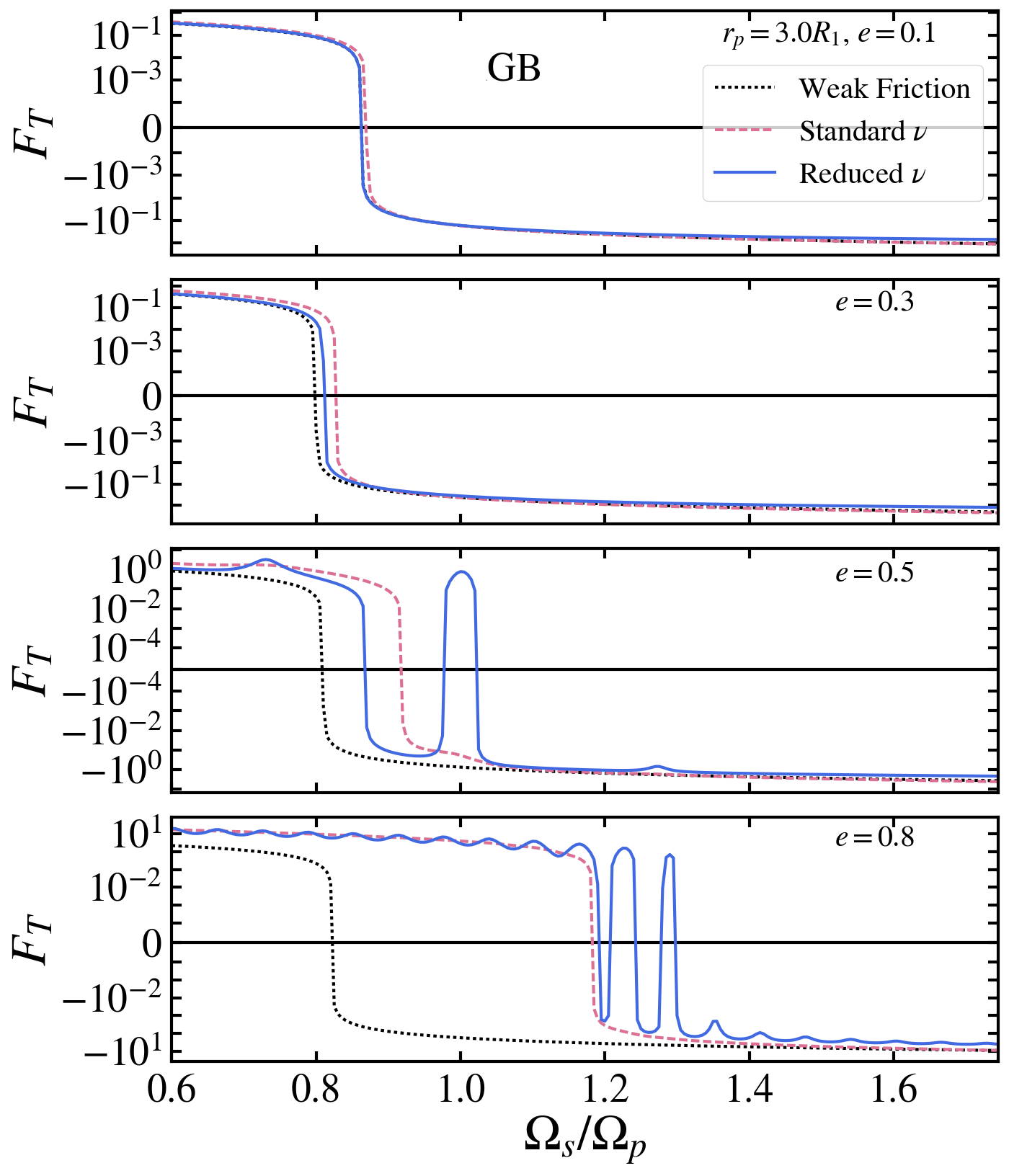}
				\caption{$F_T(e,\Omega_s/\Omega, r_p/R_1)$ (equation~\ref{eq:FT}) as a function of the rotation rate $\Omega_s$ (in units of the pericentre frequency, equation~\ref{eq:Omegap_def}) for a 10~$M_\odot$ GB stellar model depicted in Fig.~\ref{fig:StellarModels}. Results are shown for four different eccentricities, all with the same pericentre distance $r_{\rm p} = 3 R_1$. The dotted line is the weak friction result from equation~(\ref{eq:FT_WF}). The (blue) solid line uses a reduced viscosity (equation~\ref{eq:GN}) while the (red) dashed line uses the standard viscosity (equation~\ref{eq:standard}). Pseudosynchronous rotation corresponds to $F_T=0$.}
				\label{fig:FT_vs_OS_RGB}
			\end{center}
		\end{figure*}

\begin{figure*}
			\begin{center}
				\includegraphics[width = 5 in]{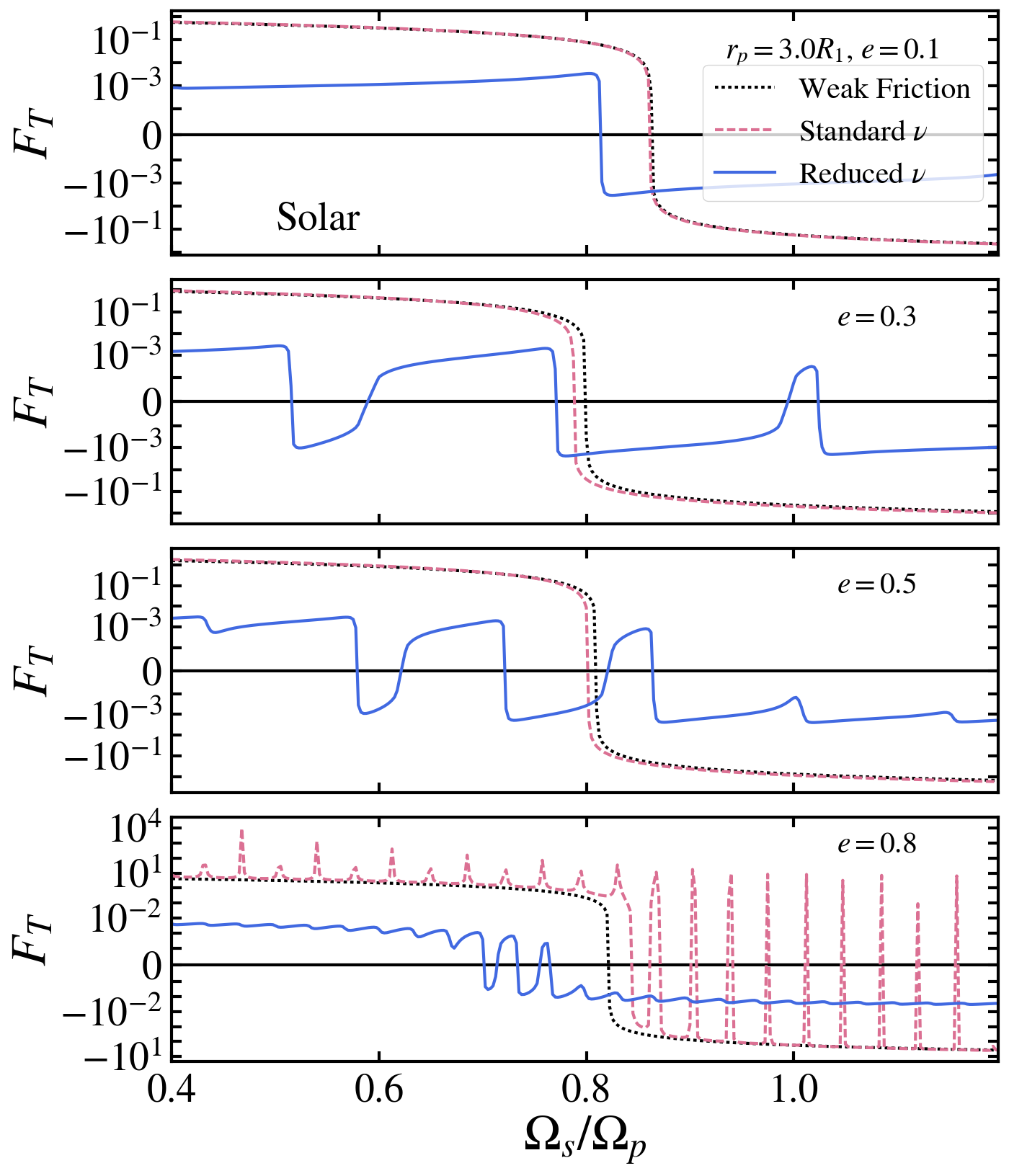}
				\caption{Same as Fig.~\ref{fig:FT_vs_OS_RGB} but for the solar-type stellar model depicted in Fig.~\ref{fig:StellarModels}.}
				\label{fig:FT_vs_OS_Solar}
			\end{center}
		\end{figure*}

	In realistic (MESA) stellar models, we use equation~(\ref{eq:FT}) to compute $F_T$ for both the standard viscosity (equation~\ref{eq:standard}) and the reduced viscosity (equation~\ref{eq:GN}). Figure~\ref{fig:FT_vs_OS_RGB} displays the results for the GB stellar model and Fig.~\ref{fig:FT_vs_OS_Solar} for the solar model. Equilibrium spin (pseudosynchronous rotation) corresponds to $F_T = 0$. For the GB model, the pseudosynchronous value of $\Omega_s$ can be nearly a factor of two larger than the predicted $\Omega_{\rm ps}$ from weak friction theory (equation~\ref{eq:WFOS}). Additionally, $F_T$ can be zero for multiple rotation rates, allowing for multiple spin equilibria (though not all of these are stable). This behavior was also noted and discussed in \citet{Storch14}. For the GB stellar model, the two viscosity prescriptions yield similar order of magnitudes for the values of $F_T$. This is unsurprising as the eddy timescale $\tau_{\rm eddy}$ is generally short throughout the convective envelope of the GB star, and the viscosity is never significantly reduced (see Fig.~\ref{fig:gammavsforcing}).
		
	For the solar-type stellar model, the result of $F_T$ with the reduced viscosity prescription is very different from either the weak friction approximation or the calculation that assumes standard (frequency-independent) viscosity (see Fig.~\ref{fig:FT_vs_OS_Solar}). In general, $|F_T|$ is 1-2 orders of magnitude smaller for the reduced viscosity than for the standard viscosity. Additionally, $F_T$ can cross zero for slower rotation rates of nearly half of $\Omega_{\rm ps}$ (equation~\ref{eq:WFOS}) when the viscosity is reduced. 
		
\begin{figure*}
			\begin{center}
				\includegraphics[width = 6.5 in]{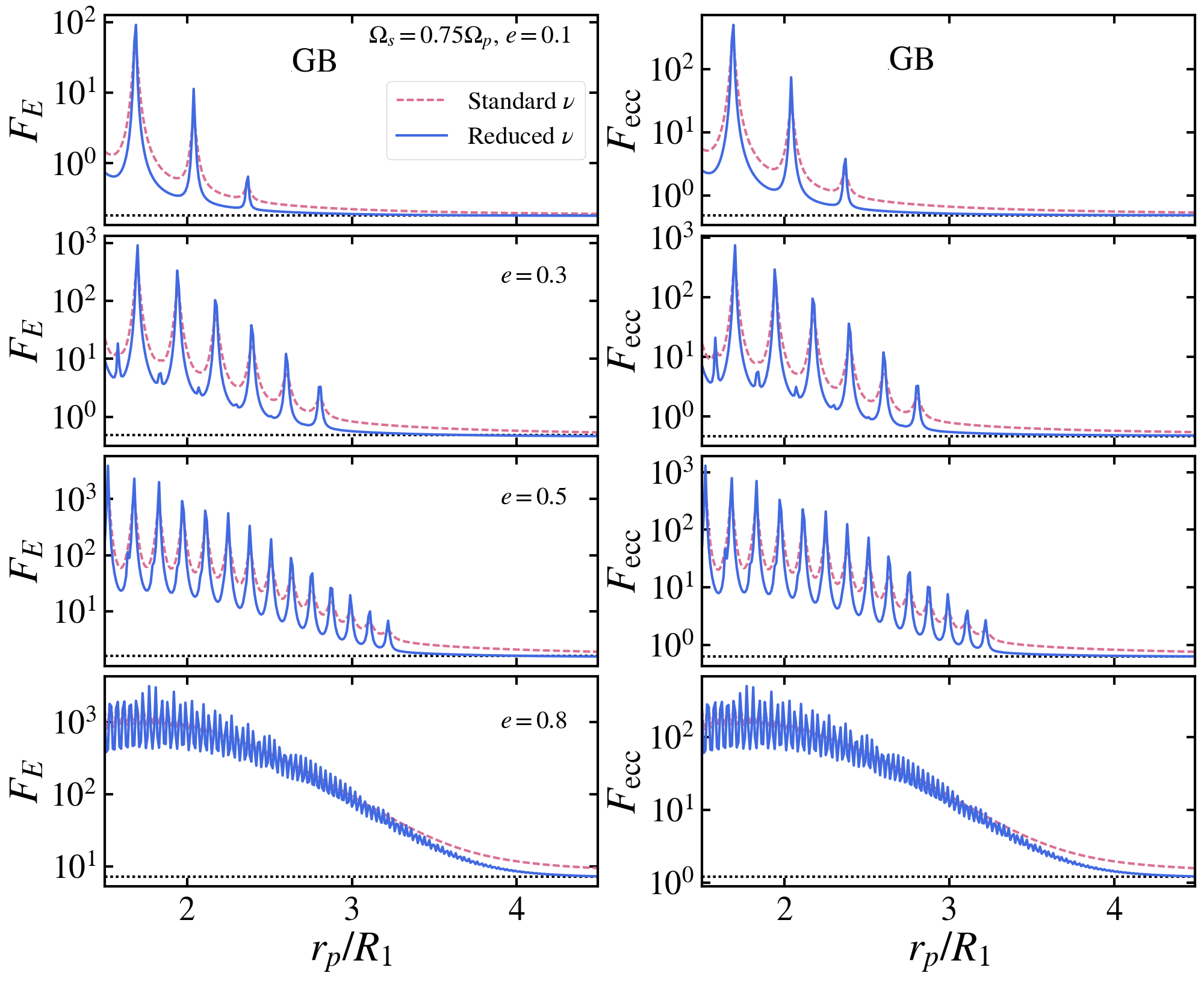}		
				\caption{$F_E(e,\Omega_s/\Omega, r_p/R_1)$ and $F_{\rm ecc}(e,\Omega_s/\Omega, r_p/R_1)$ (equations~\ref{eq:FE} and \ref{eq:FEcc}) as a function of the pericentre distance for the 10~$M_\odot$ GB stellar model depicted in Fig.~\ref{fig:StellarModels}. The stellar spin frequency is chosen to be $0.75\Omega_p$. The four pairs of panels show results for four eccentricities, as labeled. The dotted black lines correspond to the weak friction results(see equations~\ref{eq:FT_WF} and \ref{eq:FEcc_WF}). The solid lines are calculated with the reduced viscosity from equation~(\ref{eq:GN}) and the dashed lines with the frequency-independent  viscosity from equation~(\ref{eq:standard}).}
				\label{fig:FE_FT_vs_rp_RGB}
			\end{center}
		\end{figure*}

	The functions $F_E$ and $F_{\rm ecc}$ determine the orbital decay and circularization rates of the binary. Figure~\ref{fig:FE_FT_vs_rp_RGB} displays $F_E$ (left column) and $F_{\rm ecc}$ (right column) as a function of pericentre distance for the GB stellar model given a rotation rate of $\Omega_s = 0.75 \Omega_p$, slightly below $\Omega_{\rm ps}$. Each row corresponds to a different orbital eccentricity. The functions $F_E(r_p/R_1)$ and $F_{\rm ecc}(r_p/R_1)$ have strong peaks that correspond to resonances between the mode frequencies, $\omega_\alpha$, and the forcing frequencies, $\omega_{Nm}$. For higher eccentricities $(e\gtrsim 0.8)$, $F_{Nm}$ can be appreciable even for $N$ of a few times larger than $\Omega_{\rm p}/\Omega = (1+e)^{1/2}/(1-e)^{3/2}$. Thus many forcing frequencies contribute significantly to $F_E$ and $F_{\rm ecc}$ and can dominate the sum near a resonance, as seen in the bottom row of Fig.~\ref{fig:FE_FT_vs_rp_RGB}. Importantly, $F_E$ and $F_{\rm ecc}$ can be two orders of magnitude larger than the weak friction results (equations~\ref{eq:FE_WF} and \ref{eq:FEcc_WF}) for small $r_p$ and high $e$. At larger pericentre distances, $F_E$ and $F_{\rm ecc}$ agree with the weak friction results. For the GB stellar model, the choice of viscosity prescription does not have a significant effect on the calculated orbital decay and circularization rates, as expected due to the short eddy timescale in the convective zone.  
		
\begin{figure*}
			\begin{center}
				\includegraphics[width = 6.5 in]{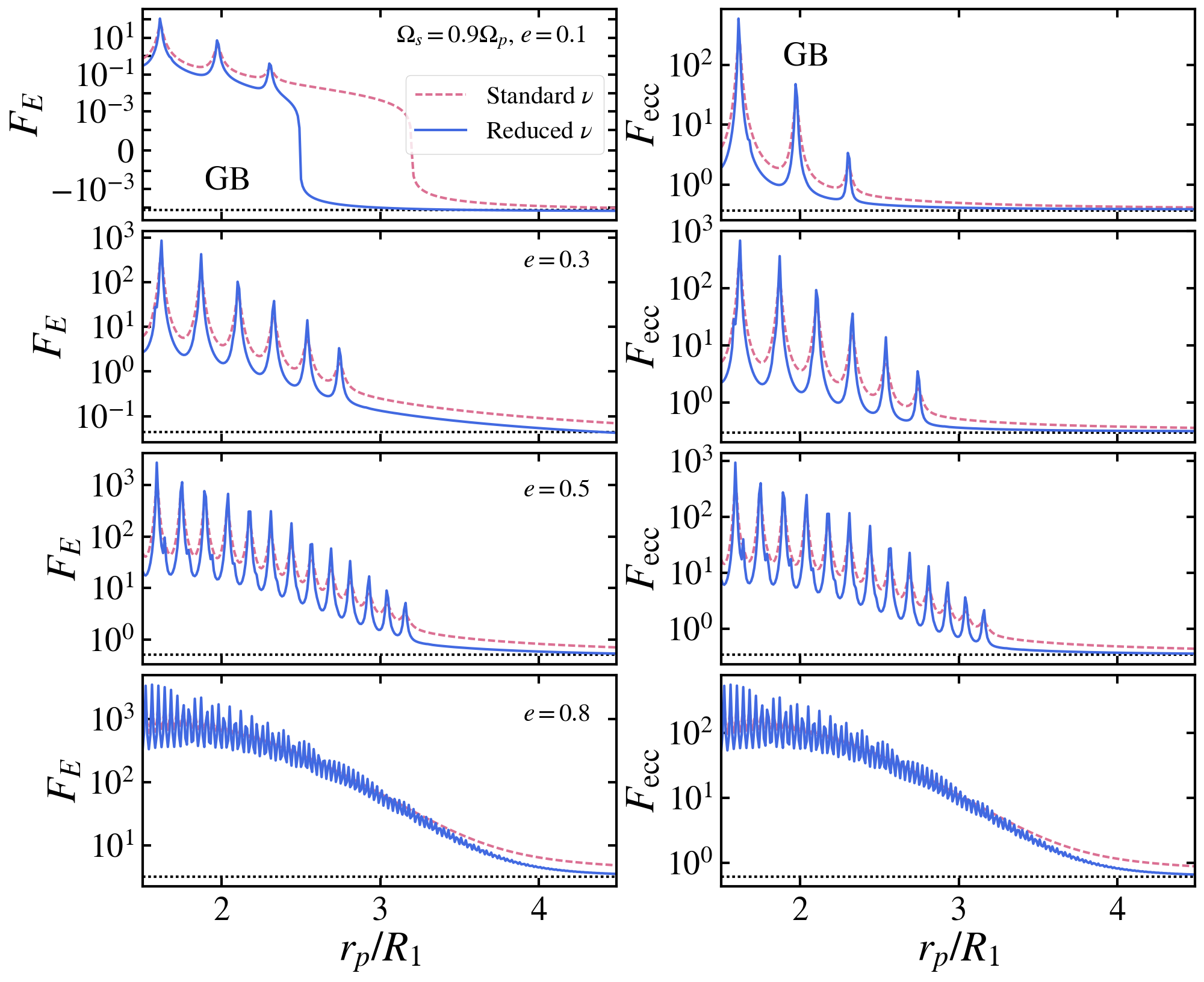}		
				\caption{Same as Fig.~\ref{fig:FE_FT_vs_rp_RGB} except the stellar spin frequency is chosen to be $0.9\Omega_p$.}
				\label{fig:FE_FT_vs_rp_RGB_OSp9}
			\end{center}
		\end{figure*}

	Figure~\ref{fig:FE_FT_vs_rp_RGB_OSp9} shows $F_E(r_p/R_1)$ and $F_{\rm ecc}(r_p/R_1)$ for the GB stellar model, as in Fig.~\ref{fig:FE_FT_vs_rp_RGB}, but for a larger spin rate of $\Omega_s=0.9 \Omega_p$. For some combinations of $e$ and $r_p$, $\Omega_s=0.9 \Omega_p$ exceeds the pseudosynchronous rotation rate, giving rise to orbital expansion ($F_E < 0$, see equation~\ref{eq:adot}). As an example, for $e=0.1$ (shown in the upper left panel of Fig.~\ref{fig:FE_FT_vs_rp_RGB_OSp9}), $F_E$ is negative for $r_p/R_1 \gtrsim 2.5-3.5$ (depending on the viscosity prescription). Otherwise, there are no qualitative differences between $F_E(r_p/R_1)$ and $F_{\rm ecc}(r_p/R_1)$ for $\Omega_s = 0.75 \Omega_p$ and $\Omega_s = 0.9 \Omega_p$.
	
\begin{figure*}
			\begin{center}
				\includegraphics[width = 6.5 in]{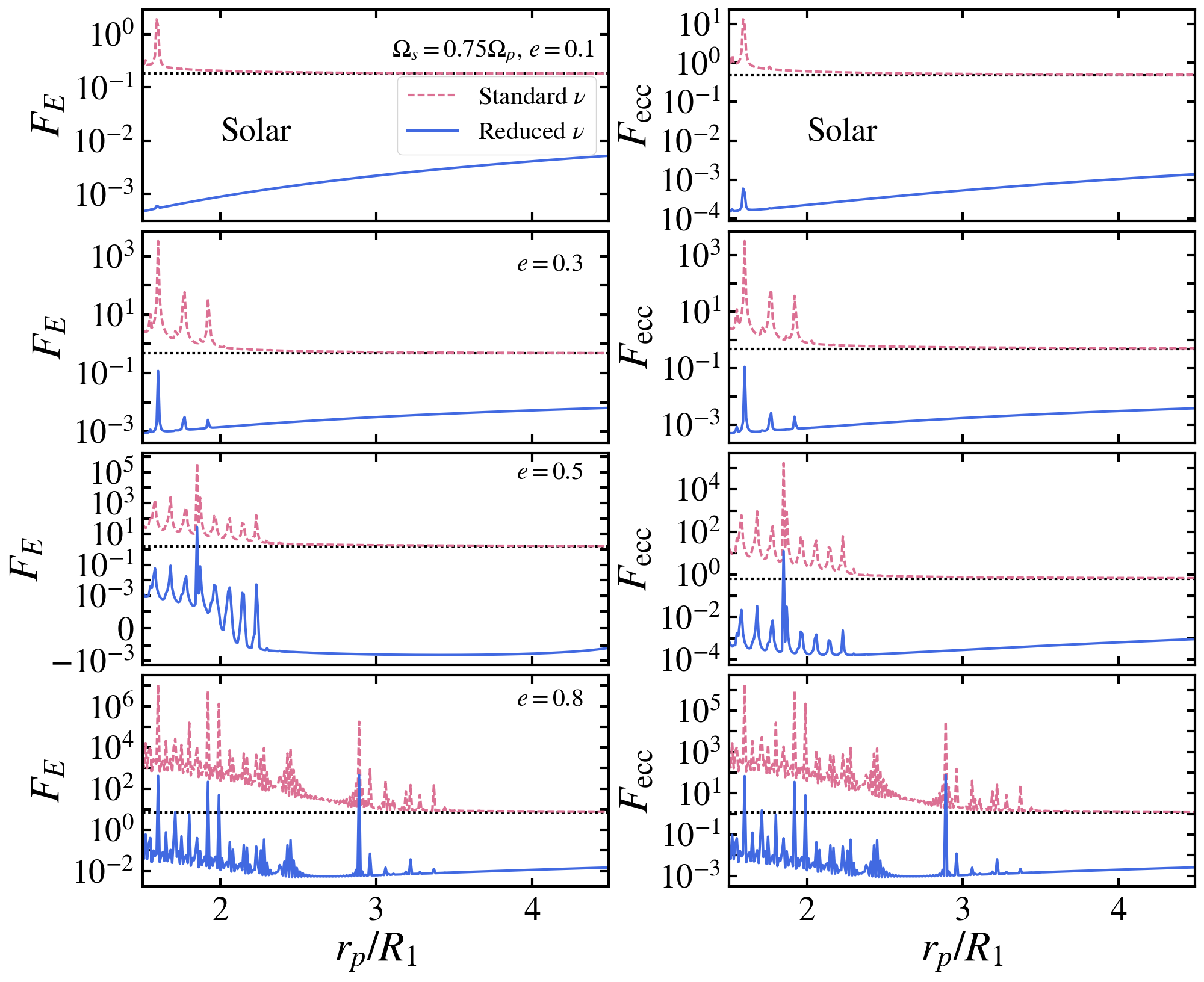}	
				\caption{Same as Fig.~\ref{fig:FE_FT_vs_rp_RGB} but for the solar-type stellar model.}
				\label{fig:FE_FT_vs_rp_Solar}
			\end{center}
		\end{figure*}

	The dimensionless orbital decay and circularization rates for the solar-type stellar model are shown in Fig.~\ref{fig:FE_FT_vs_rp_Solar}. For standard (frequency-independent) $\nu$, $F_E$ and $F_{\rm ecc}$ can be a few orders of magnitude larger than the weak friction results for small $r_p$ and high $e$ but agree with the weak friction results at larger $r_p$. Unlike for the GB model, the viscosity prescription dramatically affects the calculated $F_E$ and $F_{\rm ecc}$, evident in all panels of Fig.~\ref{fig:FE_FT_vs_rp_Solar}. Convective viscosity is inefficient in circularizing and shrinking the orbit because the eddy turnover time in the convection envelope is orders of magnitude longer than the pericentre passage time $\Omega_p^{-1}$. For a solar-type star, we expect orbital decay via convective dissipation to be a few orders of magnitude smaller than the weak friction prediction.
	
	\section{High-Eccentricity Limit: Alternative Calculation of Tidal Evolution}\label{sec:HighEcc}
	When the binary orbit is highly eccentric (with $(1-e) \ll1$), oscillation modes in the primary star are excited at pericenter and subsequently damp as the mode oscillates freely throughout the rest of the orbit. In this regime, it is possible to calculate the tidal evolution 
	in a different way \citep[cf.][]{Lai97,Fuller12a,Vick18}.
	
	A key quantity is the tidal energy transfer to a stellar mode (labeled $\alpha$) during the ``first" pericenter passage (``first" means that there is no prior oscillation in the star). This can be computed as
	\begin{equation}
	\Delta E_{\rm in,\alpha} = 2 \uppi^2 \frac{GM_2^2R_1^5}{r_{\rm p}^6} \left(\frac{\sigma_\alpha }{\epsilon_\alpha}\right)|Q_\alpha K_{2m}|^2, \label{eq:DeltaE}
	\end{equation}
	with 
	\begin{equation}
	K_{2m} = \frac{W_{2m}}{2\uppi}\int_{-P/2}^{P/2} dt\left(\frac{r_{\rm p}}{D}\right)^{3} \text{e}^{\text{i}\sigma_\alpha t - \text{i} m \Phi(t)},
	\end{equation}
	where $r_{\rm p} = a(1-e)$, $Q_\alpha$ and $\epsilon_\alpha$ are given by equations~(\ref{eq:defQ}) and (\ref{eq:defepsilon}), and $\sigma_\alpha = \omega_\alpha + m \Omega_s$ is the mode frequency in the inertial frame. Under the condition  that $\sigma_\alpha/\Omega_{\rm p} \gtrsim $ a few, the integral $K_{2m}$ can be approximated with expressions provided in Appendix C of \citet{Lai97}. Note that equation~(\ref{eq:DeltaE}) includes contributions from both the $\omega_\alpha,m$ and the physically identical $-\omega_{-\alpha},-m$ terms. The total energy transfer in a single pericentre passage is given by the restricted sum over positive-frequency modes, $\Delta E_{\rm in} = \sum_{\alpha'} \Delta E_{\rm in, \alpha}$.
	
	When the mode damping time $\Gamma_\alpha^{-1} = \Gamma^{-1}_\alpha(\omega_\alpha)$ is less than the orbital period, i.e., $\Gamma_\alpha^{-1}\lesssim P$,  the orbital energy decay rate is simply given by 
	\begin{equation}
	\dot E_{\rm orb} \simeq -\sum_{\alpha'} \frac{\Delta E_{\rm in, \alpha}}{P}.
	\end{equation}
	On the other hand, when $\Gamma_\alpha^{-1} \gtrsim P$, the orbital decay rate is \citep{Lai97, Vick18} 
	\begin{equation}
	\dot{E}_{\rm orb} = -\sum_{\alpha'} \dot{E}_{\rm in, \alpha} \simeq - 2 \sum_{\alpha'} \Gamma_\alpha E_{\rm ss,\alpha} = -2\sum_{\alpha'} \gamma_\alpha(\omega_\alpha)\frac{\omega_\alpha}{\epsilon_\alpha} E_{\rm ss,\alpha}.
	\end{equation}
	Note that the mode damping rate $\Gamma_\alpha = \Gamma_\alpha(\omega_\alpha)$ is related to $\gamma_\alpha = \gamma_\alpha(\omega_\alpha)$ (see Fig.~\ref{fig:gammavsforcing}) by $\Gamma_\alpha = \omega_\alpha \gamma_\alpha/\epsilon_\alpha$ (see Appendix). The steady-state mode energy $E_{\rm ss,\alpha}$ is given by \citep{Lai97,Fuller12a}
	\begin{equation}
	E_{\rm ss,\alpha} = \frac{\Delta E_{\rm in,\alpha}}{2\left[\cosh(\Gamma_\alpha P)-\cos(\sigma_\alpha P)\right]}. \label{eq:ESS_def}
	\end{equation}
	For a single freely oscillating mode, the tidal torque is related to the energy transfer rate in the inertial frame via
	\begin{equation}
	T_\alpha = \frac{m }{\sigma_\alpha} \dot{E}_{\rm in, \alpha}. 
	\end{equation}
	For $\Gamma_\alpha P \ll 1$ [and thus $\cosh (\Gamma_\alpha P)\simeq 1$], equation~(\ref{eq:ESS_def}) implies that a resonance occurs when $\sigma_\alpha P$ is an integer multiple of $2\uppi$. This resonance condition is the same as $\omega_\alpha = N\Omega - m \Omega_s = \omega_{Nm}$ (see equations~\ref{eq:Tdef} and \ref{eq:Edotin}).
	
	As in Section~\ref{sec:Theory}, we define the dimensionless torque and energy dissipation rates $F_T$ and $F_E$ that are related to $T$ and $\dot{E}_{\rm in}$ by equations~(\ref{eq:TGeneral}) and (\ref{eq:EdotGeneral}) respectively. When $\Gamma_\alpha P\lesssim 1,$ we have
	\begin{align}
	F_T =& \frac{5 \pi}{6}\left(\frac{\bar{\omega}_{\rm f}}{Q_{\rm f}}\right)^{2}\left(\frac{\omega_{\rm f}^2}{\gamma_{\rm f}\Omega}\right)(1+e)^6 \nonumber \\
	&\times \sum_{\alpha'}\left(\frac{\gamma_\alpha \omega_\alpha}{\epsilon_{\alpha}^2}\right)\frac{m|Q_\alpha K_{2m}|^2}{\left[\cosh(\Gamma_\alpha P)-\cos(\sigma_\alpha P)\right]} \label{eq:FT_highe}\\
	F_E =& \frac{5 \pi}{6}\left(\frac{\bar{\omega}_{\rm f}}{Q_{\rm f}}\right)^{2}\left(\frac{\omega_{\rm f}^2}{\gamma_{\rm f}\Omega}\right)(1+e)^{15/2}(1-e)^{3/2} \nonumber \\
	&\times\sum_{\alpha'}\left(\frac{\gamma_\alpha \omega_\alpha}{\epsilon_{\alpha}^2}\right)\left(\frac{ \sigma_\alpha}{\Omega}\right)\frac{|Q_\alpha K_{2m}|^2}{\left[\cosh(\Gamma_\alpha P)-\cos(\sigma_\alpha P)\right]}.\label{eq:FE_highe}
	\end{align}
	When $\Gamma_\alpha P\gtrsim1$, and energy transfer at pericenter is dissipated within a single orbit, we use
	\begin{align}
	F_T =& \frac{5}{12}\left(\frac{\bar{\omega}_{\rm f}}{Q_{\rm f}}\right)^{2}\left(\frac{\omega_{\rm f}^2}{\gamma_{\rm f}\Omega}\right)(1+e)^6\sum_{\alpha'}m\left(\frac{\Omega}{\epsilon_\alpha}\right)|Q_\alpha K_{2m}|^2 \label{eq:FT_highe_simple}\\
	F_E =& \frac{5}{12}\left(\frac{\bar{\omega}_{\rm f}}{Q_{\rm f}}\right)^{2}\left(\frac{\omega_{\rm f}^2}{\gamma_{\rm f}\Omega}\right)(1+e)^{15/2}(1-e)^{3/2} \nonumber \\
	&\times\sum_{\alpha'} \left(\frac{\sigma_\alpha}{\epsilon_\alpha}\right)|Q_\alpha K_{2m}|^2.\label{eq:FE_highe_simple}
	\end{align}
	
\begin{figure*}
			\begin{center}
				\includegraphics[width = 3.2 in]{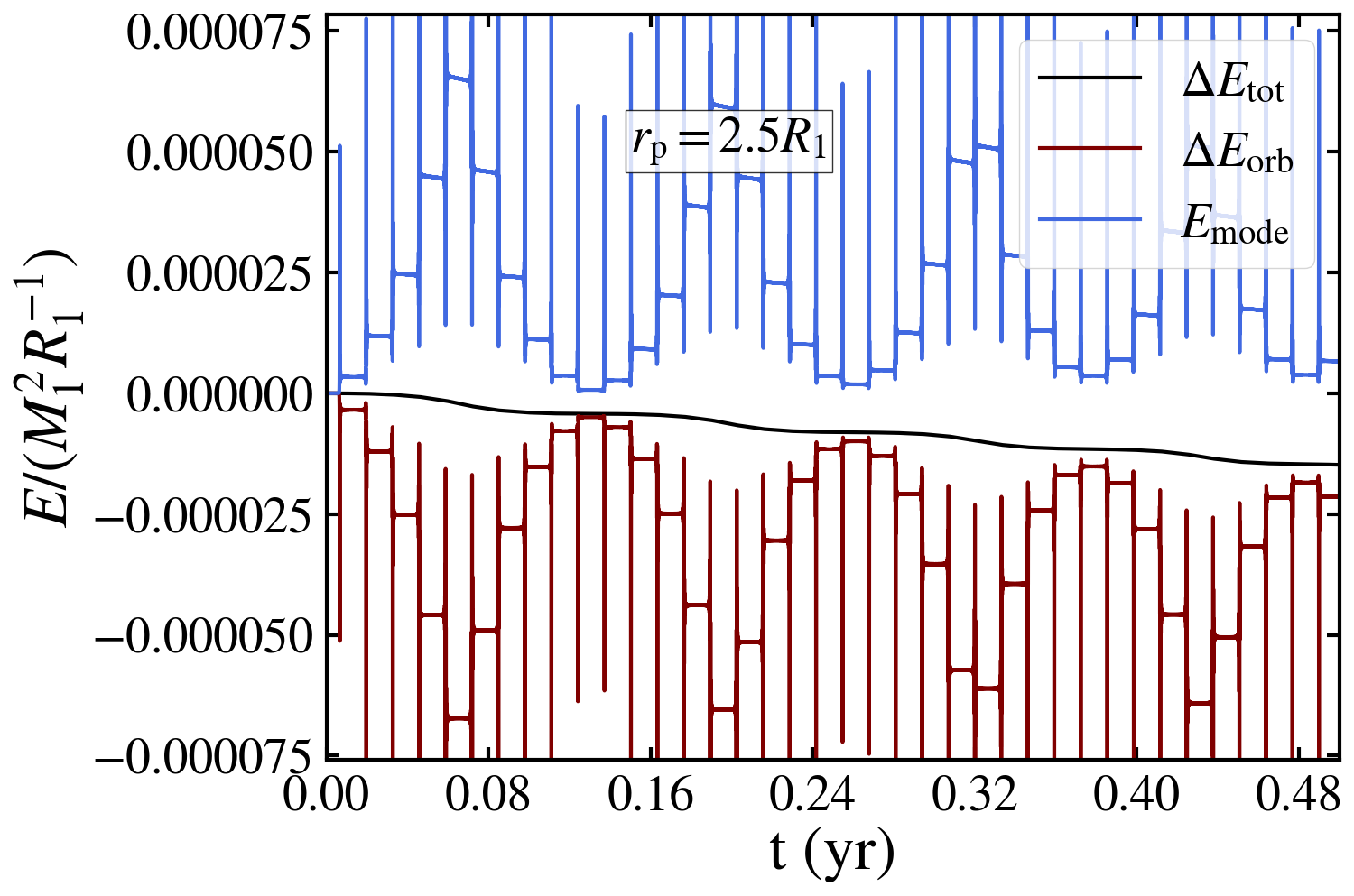}	
				\includegraphics[width = 3.2 in]{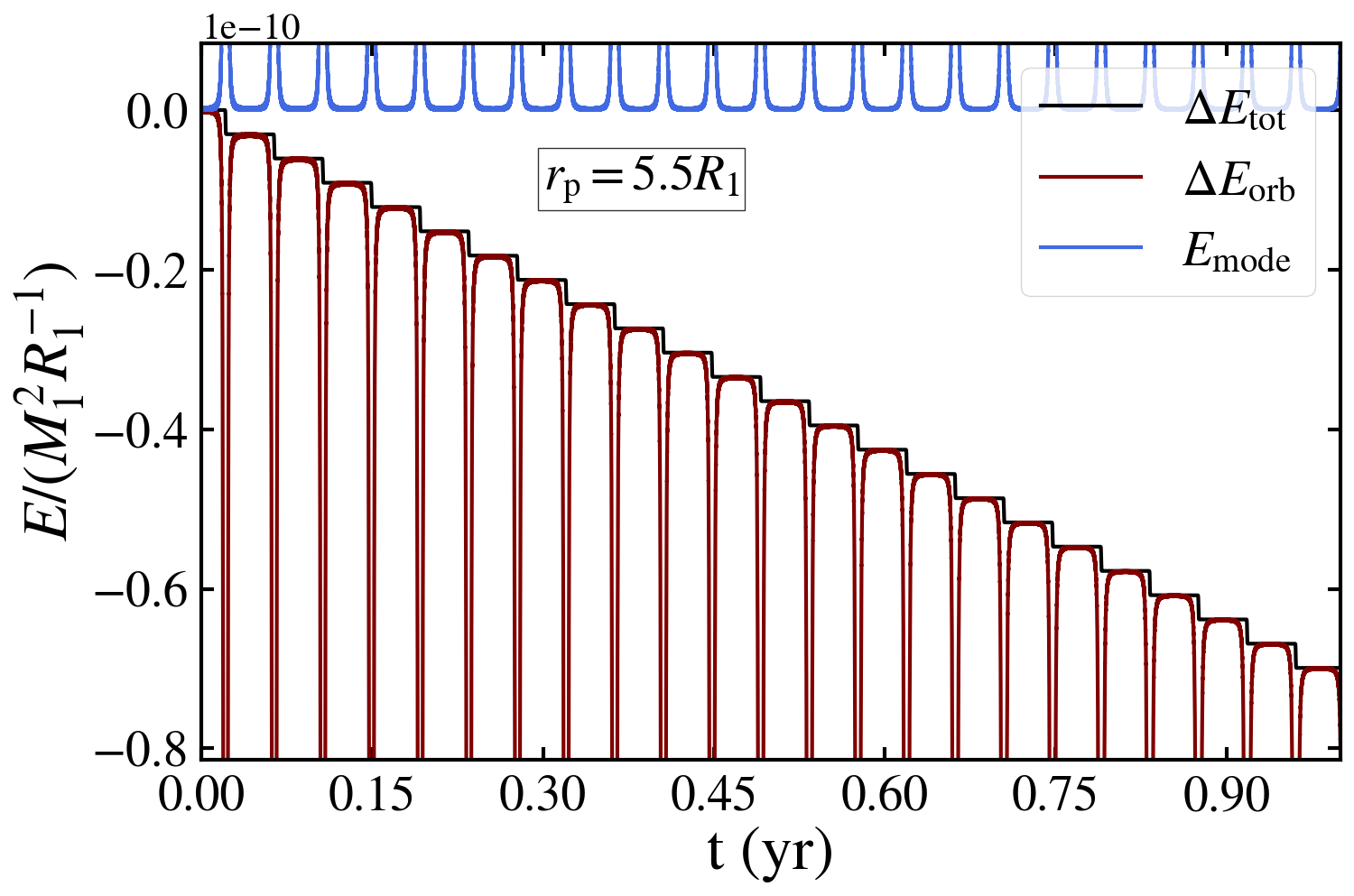}				
				\caption{The evolution of the mode energy $E_{\rm mode}$ (a sum of the energies in the $l=2,m=-2,0,2$ f-modes), orbital energy, $\Delta E_{\rm orb} = E_{\rm orb} - E_{\rm orb,0}$, and total energy $\Delta E_{\rm tot} = E_{\rm tot} - E_{\rm tot,0} = E_{\rm mode} + \Delta E_{\rm orb}$ in units where $G=M_1=R_1=1$ for a binary with the solar-type stellar model, a mass ratio of $M_2/M_1=1$, and an initial eccentricity $e = 0.85$. The strong peaks and valleys correspond to pericentre passages. In the left panel (with the initial pericentre distance $r_{\rm p}=2.5R_1$), the total energy changes smoothly, and does not show sudden changes at pericentre; for such small $r_{\rm p}$, the dissipation of forced oscillations near pericentre is negligible, and the mode energy visibly decays throughout the rest of the orbit. In the right panel (with larger $r_{\rm p}=5.5R_1$), sharp changes in the total energy at pericentre account for majority of energy dissipation in the binary.}
				\label{fig:EnergyEvolutionComparison}
			\end{center}
		\end{figure*}

	A key assumption of the above formulation of tidal evolution is that the damping of the free mode oscillations, away from pericentre, dominates the tidal dissipation rate. This is true at small $r_p$ and large eccentricity. However, as $r_p$ increases, damping of forced oscillations during pericentre passages becomes important. To illustrate this point, we carry out time-dependent calculation of the ``mode + orbit" system for an equal-mass binary with a non-rotating solar-type primary star. The secondary is treated as a point mass. Figure~\ref{fig:EnergyEvolutionComparison} shows the evolution of the energy in $l=2$ f-mode oscillations, the orbital energy, and the total energy (the sum of the mode and orbital energies), for (initial) eccentricity $e=0.85$ and two different pericentre distances ($r_p=2.5R_1$ in the left panel and $r_{\rm p} = 5.5R_1$ right panel). The coupled evolution of the mode amplitudes and the orbit was executed by combining equation~(\ref{eq:cdot}) of Section~\ref{sec:Theory} for the time evolution of the mode amplitude and equations (6) and (7) of \citet{Vick19} for the orbital evolution (with all general relativity terms set to zero). Expressions for the mode energy and orbital energy are provided in equations (12) and (13) of \citet{Vick19}. For $r_{\rm p}=2.5 R_1$, the binary is in the regime where the damping of free mode oscillations dominates the energy dissipation. In the left panel of Fig.~\ref{fig:EnergyEvolutionComparison}, we see that the total energy does not change significantly during pericentre passages, and the mode energy decays as the oscillations damp away from pericentre. In the right panel, for $r_{\rm p} = 5.5 R_1$, the dissipation of forced oscillations at pericentre dominates the energy dissipation, and the total energy of the system decreases sharply during each pericentre passage.
	
\begin{figure*}
			\begin{center}
				\includegraphics[width = 3.2 in]{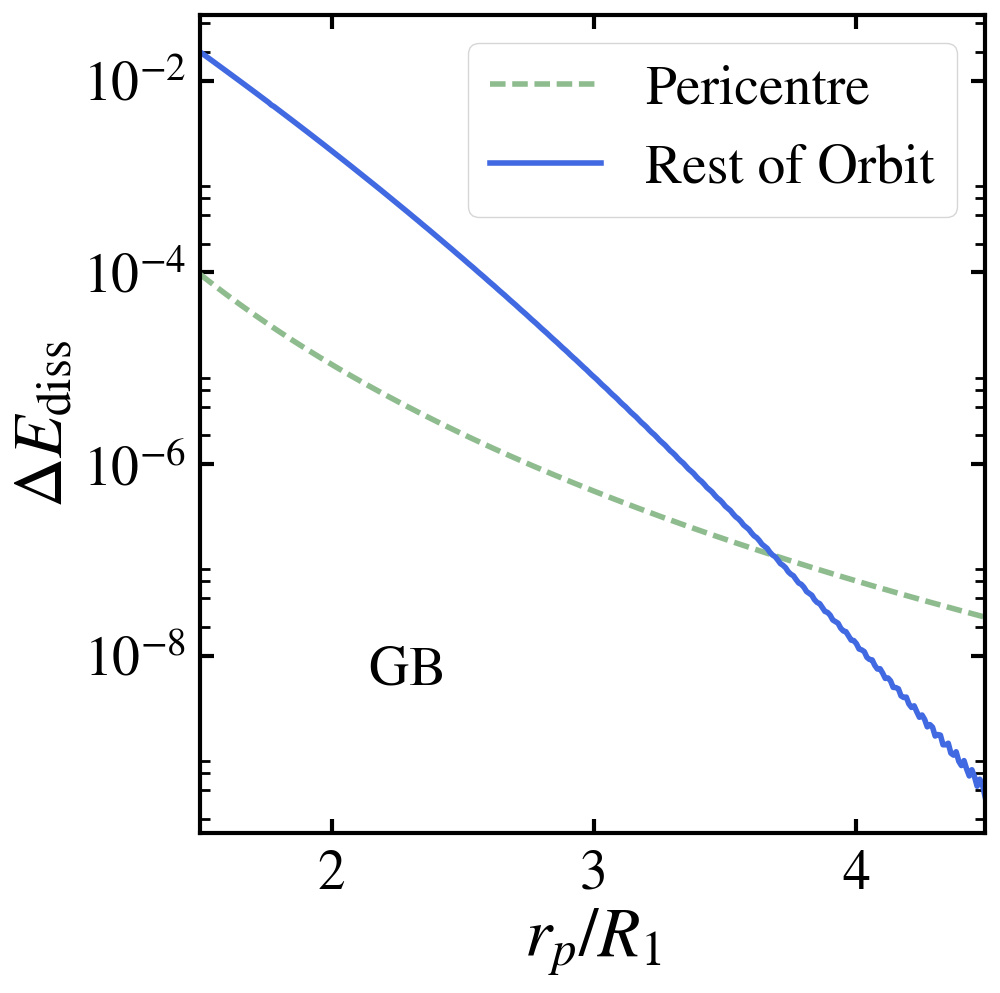}	
				\includegraphics[width = 3.2 in]{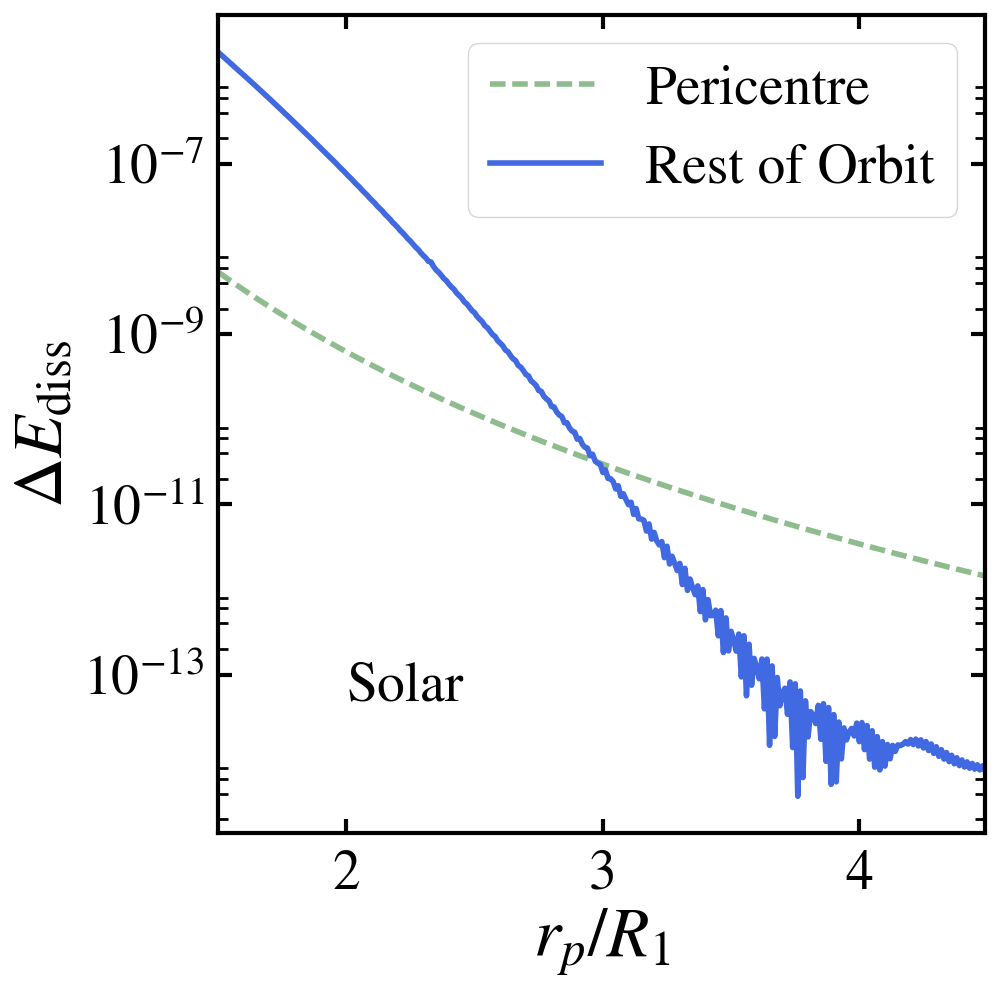}
				\caption{The energy dissipation in a single orbit near pericentre (equation~\ref{eq:Ediss_p}) and over the rest of the orbit (equation~\ref{eq:Ediss_np}) for a highly eccentric binary as a function of $r_p/R_1$ for the GB stellar model (left panel) and the solar-type model (right panel) assuming the standard (frequency-independent) viscosity. The orbital eccentricity is $e=0.85$ and the stellar rotation rate is $\Omega_s = 0.75 \Omega_p$.}
				\label{fig:Edissp_vs_Edissnp}
			\end{center}
		\end{figure*}

	We can identify the transition between the two dissipation regimes by comparing $\Delta E_{\rm diss,p}$, the amount of energy dissipated during a single pericentre passage, with $\Delta E_{\rm diss,np}$, the energy dissipated during the rest of the orbit. For simplicity, let us assume a single mode is dominant. We can estimate $\Delta E_{\rm diss,p}$ as $\gamma_\alpha(\Omega_{\rm p}) E_k / \Omega_{\rm p}$,  (see equation~\ref{eq:Edot_alphaN}), where $\Omega_{\rm p}$ is the orbital frequency at pericentre (equation~\ref{eq:Omegap_def}), and $E_k$ is the kinetic energy in the oscillations at pericentre, given by $E_k \sim k_2 M_1 (R_1 \epsilon_p)^2 \Omega_p^2/\bar{\omega}_{\rm f}^2$, with $\epsilon_p = M_2 R_1^3/(M_1 r_p^3)$. Then,
	\begin{equation}
	\Delta E_{\rm diss,p} \sim \gamma_\alpha(\Omega_p) \Omega_p M_1 R_1^2 \left(\frac{k_2}{\bar{\omega}_{\rm f}^2}\right)\left(\frac{M_2}{M_1}\right)^2\left(\frac{R_1}{r_p}\right)^6.\label{eq:Ediss_p}
	\end{equation}
	The energy dissipated in a single orbit away from pericentre is 
	\begin{equation}
	\Delta E_{\rm diss,np} \sim \min[1,\Gamma_\alpha(\omega_\alpha)P] \Delta E_{\rm in, \alpha},\label{eq:Ediss_np}
	\end{equation}
	where we have neglected resonances (which occur at $\sigma_\alpha = N \Omega$), and $\Delta E_{\rm in, \alpha}$ is given by equation~(\ref{eq:DeltaE}). In Fig.~\ref{fig:Edissp_vs_Edissnp}, we compare equations~(\ref{eq:Ediss_p}) and (\ref{eq:Ediss_np}) for both the GB and solar-type stellar models assuming the standard (frequency-independent) viscosity. We find that $E_{\rm diss,p} \gtrsim E_{\rm diss,np}$ (i.e. tidal energy dissipation occurs primarily during the pericentre passage) when $r_p \gtrsim 3.7 R_1$ for the GB star and $r_p \gtrsim 3 R_1$ for the solar-type star. Therefore, equations~(\ref{eq:FT_highe})-(\ref{eq:FE_highe_simple}) are only accurate for small $r_p$ and we expect deviation from the results of Section~\ref{sec:Results} when $r_p \gtrsim 3 R_1$.
	
\begin{figure*}
			\begin{center}
				\includegraphics[width = 6.5 in]{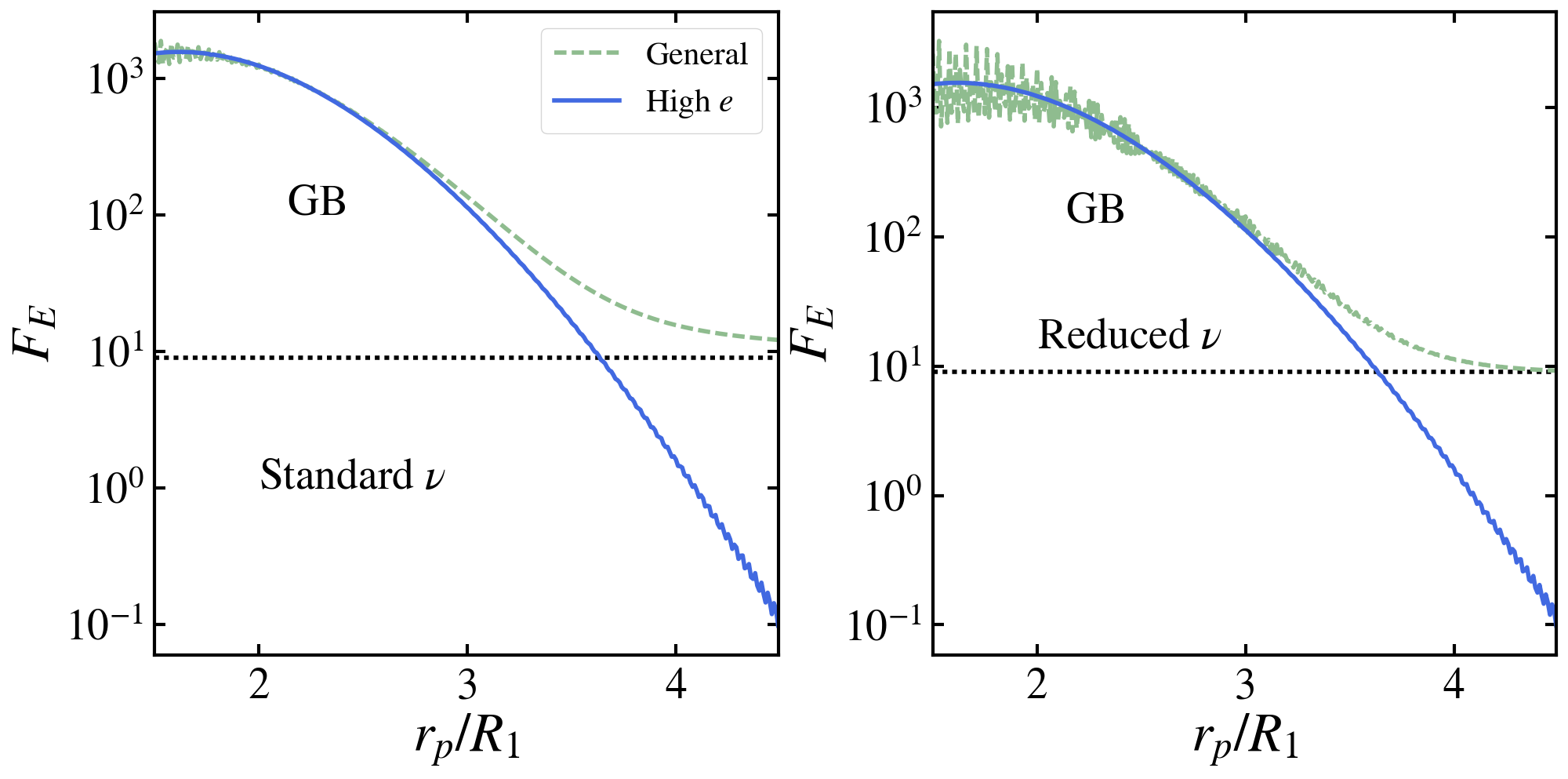}
				\caption{The dimensionless tidal energy transfer rate $F_E(e,\Omega_s/\Omega, r_p/R_1)$ (equations~\ref{eq:FE}) vs the pericentre distance for the 10~$M_\odot$ GB stellar model. The orbital eccentricity is $e=0.85$ and the stellar rotation rate is $\Omega_s = 0.75 \Omega_p$. The left panel is calculated with the standard (frequency-independent) viscosity and the right panel with the reduced viscosity from equation~(\ref{eq:GN}). The solid lines are obtained using the high-eccentricity expression (equations~\ref{eq:FE_highe} and \ref{eq:FE_highe_simple}), and the dashed green lines are obtained using the general expression (equation~\ref{eq:FE}). The dotted black lines correspond to the weak friction result (equation~\ref{eq:FE_WF}).}
				\label{fig:RGB_FE_vs_rp_highe}
			\end{center}
		\end{figure*}

\begin{figure*}
			\begin{center}
				\includegraphics[width = 6.5 in]{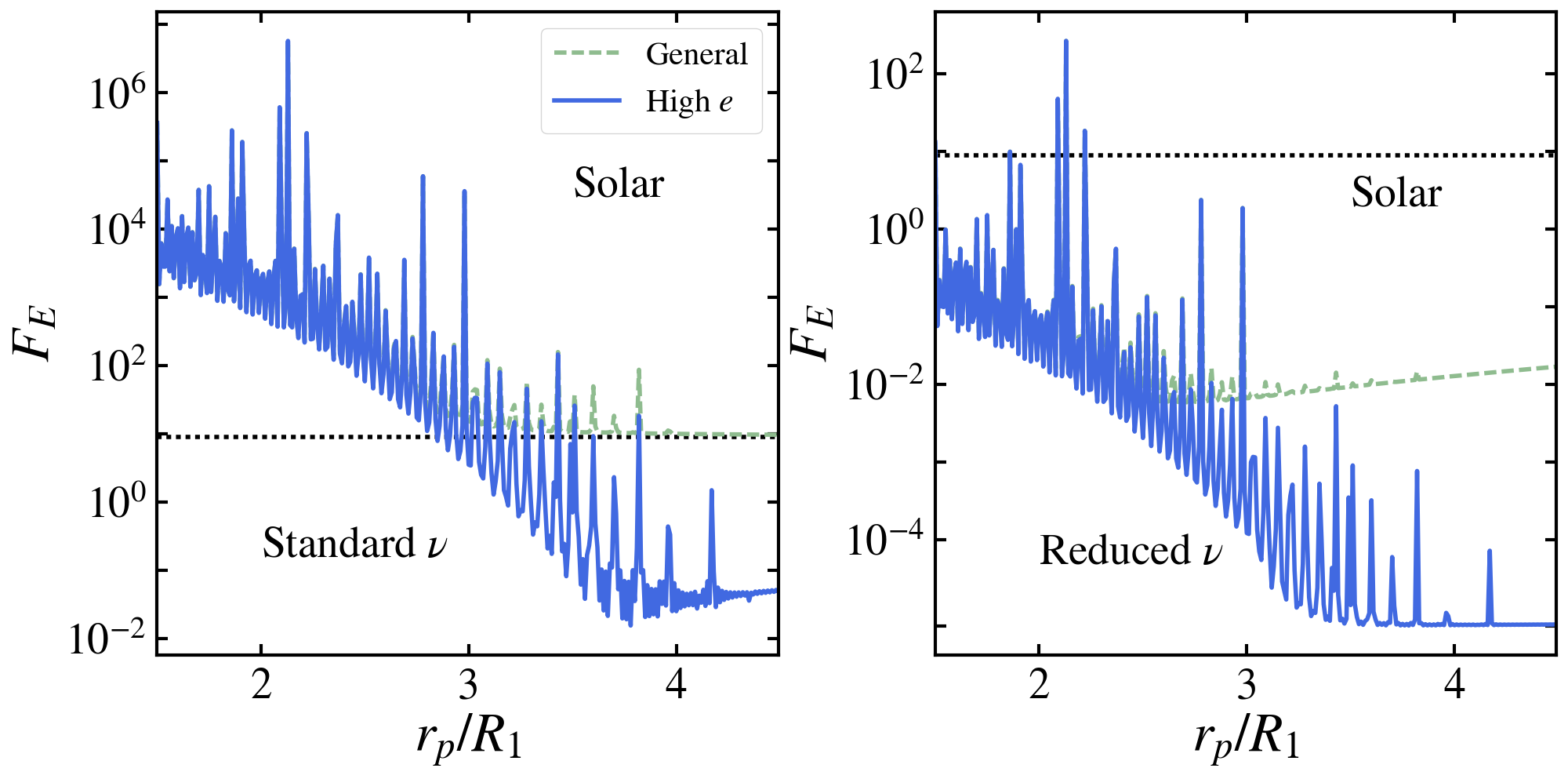}				
				\caption{Same as Fig.~\ref{fig:RGB_FE_vs_rp_highe} but for the solar-type stellar model. Note that in both panels, the high-eccentricity result (equation~\ref{eq:FE_highe}) is in near perfect agreement with the general expression (equation~\ref{eq:FE}) for small $r_{\rm p}$.}
				\label{fig:Solar_FE_vs_rp_highe}
			\end{center}
		\end{figure*}

	In Figs.~\ref{fig:RGB_FE_vs_rp_highe} and \ref{fig:Solar_FE_vs_rp_highe}, we show the dimensionless energy transfer rate for a highly eccentric binary (equation~\ref{eq:FE_highe}) as a function of $r_p/R_1$  and compare with the general expression from equation~(\ref{eq:FE}). For the GB model, the mode damping time, $\Gamma^{-1}$, is shorter than the orbital period for the parameters covered in Fig.~\ref{fig:RGB_FE_vs_rp_highe} [$\Gamma_\alpha^{-1} = 4.5~\text{yr} \times (\omega_\alpha/\epsilon_\alpha)$ using the standard viscosity while the orbital period is $P=12.8~\text{yr} \times (M_1/M_t)^{1/2}(r_{\rm p}/R_1)^{3/2}$ for the chosen eccentricity of $e=0.85$ and total mass $M_t = M_1+M_2$]. Therefore $F_E$ is given by equation~(\ref{eq:FE_highe_simple}), and the solid lines in the left and right panels are identical. For $r_p 
	\lesssim 3 R_1$, the general expression for $F_E$ (equation~\ref{eq:FE}) agrees with the high-eccentricity calculation (ignoring peaks due to resonances between a mode and a component of the tidal forcing). Note that the derivation of equation~(\ref{eq:FE}) assumes that the mode damping rate is longer than an orbital period, so the high-eccentricity calculation (equation~\ref{eq:FE_highe_simple}) should be more accurate in this regime. For $r_p \gtrsim 3 R_1$, the high-eccentricity expression no longer agrees with equation~(\ref{eq:FE}) and the weak friction result, as expected, because dissipation near the pericentre, where the oscillation modes are strongly forced, becomes the dominant contributor to the energy and angular momentum transfer rates. 
	
	Figure~\ref{fig:Solar_FE_vs_rp_highe} displays $F_E$ for the solar-type stellar model. Here, the viscous damping time, $\Gamma_\alpha(\omega_\alpha)^{-1}$, is much longer than the orbital period [$\Gamma_\alpha^{-1} = 1.5~\text{yr}\times(\omega_\alpha/\epsilon_\alpha)$ using the standard viscosity and $P=4.7\times10^{-3}~\text{yr}\times(M_1/M_t)^{1/2}(r_{\rm p}/R_1)^{3/2}$ for $e=0.85$], so equation~(\ref{eq:FE_highe}) is appropriate for the high-eccentricity regime. The function $F_E(r_p/R_1)$ has strong peaks that correspond to resonances between the orbital frequency and the mode frequency in the inertial frame. From Fig.~\ref{fig:Solar_FE_vs_rp_highe}, we see that the high-eccentricity calculation agrees well with the general calculation from equation~(\ref{eq:FE}) at small $r_{\rm p}$ for both the standard (left panel) and the reduced (right panel) viscosity. As with the GB model, the high-eccentricity prescription under-predicts the dissipation rates for larger $r_p$ because it does not include mode damping near the pericentre.
	
\section{Summary and Discussion} \label{sec:Discussion}
We have developed a general formalism for calculating the orbital decay and circularization rates as well as the spin synchronization rate of a star with a convective envelope in an eccentric binary. Our formalism allows for frequency-dependent turbulent viscosity reduction, which is important in the convective envelope when the eddy turnover time is longer than the tidal forcing period. The most general results are summarized in equations~(\ref{eq:FT})-(\ref{eq:td_def}). In the slow-rotation limit, and assuming that the mode frequency is much larger than the forcing frequencies, and that the damping time is independent of the forcing frequency, these general expressions reduce to the well-known weak friction results. 

In Section~\ref{sec:Results}, we calculated the dimensionless orbital decay, circularization and spin evolution rates for both GB and solar-type stellar models.
We found that the pseudosynchronous rotation rate of the star can be almost a factor of two faster than the weak tidal friction prediction for a GB star (Fig.~\ref{fig:FT_vs_OS_RGB}) and a factor of a few slower for a solar-type star (Fig.~\ref{fig:FT_vs_OS_Solar}).
We also found that, at small pericentre distances and high eccentricities, the orbital decay and circularization rates can be a few orders of magnitude larger than the prediction from weak tidal friction for an GB star (where the eddy turnover time is fast enough that viscosity reduction is negligible) and a few orders of magnitude smaller for a solar-type star due to viscosity reduction (see Figs.~\ref{fig:FE_FT_vs_rp_RGB}, \ref{fig:FE_FT_vs_rp_RGB_OSp9}, and \ref{fig:FE_FT_vs_rp_Solar}).

Lastly, in Section~\ref{sec:HighEcc} we presented a simpler calculation of the dissipation rates for highly eccentric orbits that only requires a sum over oscillation modes (rather than a sum over both the oscillation modes and many forcing frequencies). The key results are summarized in equations~(\ref{eq:FT_highe})-(\ref{eq:FE_highe_simple}) (in conjunction with equations~\ref{eq:TGeneral} and \ref{eq:EdotGeneral}). This approach neglects dissipation near pericenter and is valid for $r_{\rm p} \lesssim 3 R_1$. 

Our results are relevant to understanding populations of binary systems with evolved stars \citep[e.g.][]{Shporer16}, and those with solar-type stars, such as some of the Kepler Heartbeat stars \citep{PW18} and systems containing close-in giant planets. Our general equations can be used to track the spin and orbital evolution of a star in a binary system to answer questions such as how often binary systems will retain eccentricity at the onset of a common envelope phase and to reassess the importance of turbulent viscosity in the host star in the orbital decay of a giant planet. 

One intriguing behavior that we did not discuss in this paper is resonance-locking. In this scenario, a system encounters a resonance between the frequency of a stellar oscillation mode and the orbital frequency. As the stellar spin and orbit (and perhaps the stellar structure) evolve, the mode frequency and orbital frequency change in lock-step, maintaining the resonance \citep{Witte99, Fuller12a}. \citet{IP04b} identified the possibility of a similar behavior where significant viscosity reduction in the primary could cause an eccentric binary to evolve through multiple resonances between the primary star's rotation rate and the orbital frequency. Orbital decay is significantly enhanced while a resonance persists (see Figs.~\ref{fig:FE_FT_vs_rp_RGB},\ref{fig:FE_FT_vs_rp_RGB_OSp9}, and \ref{fig:FE_FT_vs_rp_Solar}). In some binary stellar systems, resonance-locking may set the timescale for orbital decay. 

\section*{Acknowledgements} 
We thank Morgan MacLeod for useful discussion and for motivating us to undertake this study. This work has been supported in part by NASA grants
NNX14AG94G and 80NSSC19K0444, and NSF grant
AST-17152. MV is supported by a NASA Earth and Space
Sciences Fellowship in Astrophysics. 


\appendix
\section{The Relationship between Damping Rates}
We have assumed that viscous dissipation is solely responsible for the damping of oscillation modes. In equation~(\ref{eq:defgamma}), we defined the damping rate $\gamma_\alpha(\omega_{Nm})$ by relating the viscous dissipation rate of mode $\alpha$ oscillating at the frequency $\omega_{Nm}$ to the kinetic energy of the mode. However, this damping rate is different from $\Gamma_\alpha(\omega_{Nm})$ (see equation~\ref{eq:cdot}), which relates the energy dissipation rate to the total energy of the mode. Here, we derive the relationship between the two damping rates.

The total dissipation rate in the rotating frame is equal to the tidal energy transfer rate in the same frame, given by
\begin{equation}
\dot{E} = \int d^3 x \rho \frac{\partial \bxi^*}{\partial t} \cdot (- \nabla U).
\end{equation}
Decomposed into a sum over oscillation modes and forcing frequencies (see equations~\ref{eq:potentialSum}, \ref{eq:modedef}, and \ref{eq:calphaN_def}), this is
\begin{equation}
\dot{E} = \int d^3 x \rho \sum_{\alpha N} \sum_{N' m'}\dot{c}^*_{\alpha N} U_{N'm'} \text{e}^{-\text{i}\omega_{N'm'} t} \bxi^*_\alpha(\textbf{r}) \cdot \nabla(r^2 Y_{2m'}).
\end{equation}
Using equations~(\ref{eq:UNm}),(\ref{eq:defQ}), and (\ref{eq:c_solution}), and averaging over time, the energy dissipation rate is 
\begin{align}
\dot{E} =& M_1 R_1^2 \sum_{\alpha N} \frac{\omega_{Nm}}{2 \epsilon_\alpha} \frac{(U_{Nm} Q_\alpha)^2 \Gamma_\alpha(\omega_{Nm})}{\left[(\omega_\alpha - \omega_{Nm})^2 + \Gamma^2_\alpha(\omega_{Nm})\right]} \nonumber \\
=& 2 M_1 R_1^2 \sum_{\alpha N} \Gamma_\alpha(\omega_{Nm}) \omega_{Nm} \epsilon_\alpha |c_{\alpha N}|^2. 
\end{align}
The above implies that the energy dissipation rate associated with each mode and forcing frequency $\omega_{Nm}$ is
\begin{equation}
\dot{E}_{\alpha N} = 2 \Gamma_\alpha(\omega_{Nm}) \omega_{Nm} \epsilon_\alpha |c_{\alpha N}|^2 M_1 R_1^2. \label{eq:Edot_alphaN}
\end{equation}
By comparing equations~(\ref{eq:defgamma}) and (\ref{eq:Edot_alphaN}), we obtain a relationship between $\gamma_\alpha (\omega_{Nm})$ and $\Gamma_\alpha (\omega_{Nm})$: 
\begin{equation}
\Gamma_\alpha (\omega_{Nm}) = \gamma_\alpha(\omega_{Nm}) \frac{\omega_{Nm}}{\epsilon_\alpha}.
\end{equation}
Similarly, for a freely oscillating mode, $\Gamma_\alpha(\omega_\alpha) = \gamma_\alpha(\omega_\alpha) \omega_\alpha/\epsilon_\alpha$.

\bibliographystyle{mnras}	
\bibliography{References}

\bsp
\label{lastpage}

\end{document}